\documentclass[lettersize,journal]{IEEEtran}
\usepackage{amsmath,amsfonts,color,bm}
\usepackage{algorithmic}
\usepackage{algorithm}
\usepackage{array}
\usepackage{textcomp}
\usepackage{stfloats}
\usepackage{url}
\usepackage{verbatim}
\usepackage{cite}
\usepackage{subfigure,graphicx, epstopdf}

\DeclareMathAlphabet{\eurm}{U}{eur}{m}{n}
\DeclareMathAlphabet{\mathbsf}{OT1}{cmss}{bx}{n}
\DeclareMathAlphabet{\mathssf}{OT1}{cmss}{m}{sl}
\DeclareMathAlphabet{\mathcsf}{OT1}{cmss}{sbc}{n}

\newcommand{\randomvalue}[1]{\eurm{\uppercase{#1}}}

\DeclareSymbolFont{bsfletters}{OT1}{cmss}{bx}{n}  
\DeclareSymbolFont{ssfletters}{OT1}{cmss}{m}{n}
\DeclareMathSymbol{\bsfGamma}{0}{bsfletters}{'000}
\DeclareMathSymbol{\ssfGamma}{0}{ssfletters}{'000}
\DeclareMathSymbol{\bsfDelta}{0}{bsfletters}{'001}
\DeclareMathSymbol{\ssfDelta}{0}{ssfletters}{'001}
\DeclareMathSymbol{\bsfTheta}{0}{bsfletters}{'002}
\DeclareMathSymbol{\ssfTheta}{0}{ssfletters}{'002}
\DeclareMathSymbol{\bsfLambda}{0}{bsfletters}{'003}
\DeclareMathSymbol{\ssfLambda}{0}{ssfletters}{'003}
\DeclareMathSymbol{\bsfXi}{0}{bsfletters}{'004}
\DeclareMathSymbol{\ssfXi}{0}{ssfletters}{'004}
\DeclareMathSymbol{\bsfPi}{0}{bsfletters}{'005}
\DeclareMathSymbol{\ssfPi}{0}{ssfletters}{'005}
\DeclareMathSymbol{\bsfSigma}{0}{bsfletters}{'006}
\DeclareMathSymbol{\ssfSigma}{0}{ssfletters}{'006}
\DeclareMathSymbol{\bsfUpsilon}{0}{bsfletters}{'007}
\DeclareMathSymbol{\ssfUpsilon}{0}{ssfletters}{'007}
\DeclareMathSymbol{\bsfPhi}{0}{bsfletters}{'010}
\DeclareMathSymbol{\ssfPhi}{0}{ssfletters}{'010}
\DeclareMathSymbol{\bsfPsi}{0}{bsfletters}{'011}
\DeclareMathSymbol{\ssfPsi}{0}{ssfletters}{'011}
\DeclareMathSymbol{\bsfOmega}{0}{bsfletters}{'012}
\DeclareMathSymbol{\ssfOmega}{0}{ssfletters}{'012}

\newcommand{\rvX}{{\randomvalue{X}}}  	

\newcommand{\rvs}{{\randomvalue{s}}}	

\newcommand{\rvu}{{\randomvalue{u}}}	

\newcommand{\rvv}{{\randomvalue{v}}}	

\newcommand{\rvw}{{\randomvalue{w}}}	

\newcommand{\rvx}{{\randomvalue{x}}}	

\newcommand{\rvy}{{\randomvalue{y}}}	

\newcommand{\rvz}{{\randomvalue{z}}}	

\newtheorem{thm}{Theorem}
\newtheorem{cor}{Corollary}

\newtheorem{rem}{Remark}

\definecolor{purple}{RGB}{169,87,162}

\begin{document}

\title{\huge Generalized Nearest Neighbor Decoding: General Input Constellation and a Case Study of Interference Suppression}

\author{Shuqin Pang and  Wenyi Zhang,~\IEEEmembership{Senior Member, IEEE}
        
\thanks{This work was supported in part by the National Natural Science Foundation of China under Grant 62231022.
	
	The authors are with the  Department of Electronic Engineering and Information Science, University of Science and Technology of China, Hefei 230027,  China (e-mail: shuqinpa@mail.ustc.edu.cn;  wenyizha@ustc.edu.cn).}}



\maketitle

\begin{abstract}
In this work, generalized nearest neighbor decoding (GNND), a recently proposed receiver architecture, is studied for channels under general input constellations, and multiuser uplink interference suppression is employed as a case study for demonstrating its potential. In essence, GNND generalizes the well-known nearest neighbor decoding, by introducing a symbol-level memoryless processing step, which can be rendered seamlessly compatible with Gaussian channel-based decoders. First, criteria of the optimal GNND are derived for general input constellations, expressed in the form of conditional moments matching, thereby generalizing the prior work which has been confined to Gaussian input. 
Then, the optimal GNND is applied to the use case of multiuser uplink, for which the optimal GNND is shown to be capable of achieving information rates nearly identical to the channel mutual information. By contrast, the commonly used channel linearization (CL) approach incurs a noticeable rate loss. A coded modulation scheme is subsequently developed, aiming at implementing GNND using off-the-shelf channel codes, without requiring iterative message passing between demodulator and decoder. Through numerical experiments it is validated that the developed scheme significantly outperforms the CL-based scheme. 
\end{abstract}


\section{Introduction}
\label{sec:intro}
\IEEEPARstart{I}{n}  Shannon's article ``Communication in the presence of noise'' \cite{shannon49:ire}, the now famous capacity formula of additive white Gaussian noise (AWGN) channels was established via a geometric argument. A key insight therein is that optimal decoding in AWGN channels is equivalent to searching for the nearest neighbor in a high-dimensional Euclidean space, i.e., given a received signal vector, decoding it as the codeword that has the smallest Euclidean distance to it. This geometric view has been made popular in a series of influential works (see, e.g., the widely used textbook \cite{wozencraft66:book}), and has been instrumental to understanding and implementation of digital communication systems, with a long-lasting impact till today.

The nearest neighbor decoding principle is optimal, i.e., satisfying the maximum likelihood (ML) principle, only when the channel is AWGN or conditionally AWGN (e.g., fading channels with fading coefficients perfectly known at the decoder). {The implementation of ML decoding is often prohibitively complex in general channels. Attributed to simplicity and robustness of the nearest neighbor decoding principle, even for scenarios where it loses its optimality, it has still been extensively adopted. In doing so, channel linearization (CL) is usually a necessary step rendering the channel to behave like an additive noise channel.}


A general form of CL is as follows. Consider a memoryless channel with input $\rvx \in \mathcal{X} \subseteq \mathbb{C}$, output $\rvy \in \mathcal{Y}$, state $\rvs \in \mathcal{S}$, and receive side information $\rvv \in \mathcal{V}$, satisfying the Markov chain $\rvv \leftrightarrow \rvs \leftrightarrow (\rvx, \rvy)$.\footnote{Such a Markov chain holds when the receive side information $\rvv$ is obtained via a separate measurement independent of the channel transmission from $\rvx$ to $\rvy$; for example, in a fading channel, $\rvv$ can be the received pilot at the decoder.} {The CL receiver architecture essentially decomposes the channel output into  the sum of a scaled version of the channel input  and an uncorrelated residual term, expressed as (see, e.g., \cite[Eqn. (89)]{wang22it})}
\begin{equation}
    \label{eqn:channel-linearization}
    \rvy = \frac{\mathbf{E}\left[\rvx^\dag \rvy | v\right]}{P} \rvx + \rvw(v), 
\end{equation}
conditioned upon $\rvv = v$. This way, the ``noise part'' $\rvw(v)$ is rendered conditionally uncorrelated with the ``signal part'' $\frac{\mathbf{E}\left[\rvx^\dag \rvy | v\right]}{P} \rvx$, and the decoder simply treats (\ref{eqn:channel-linearization}) as a conditionally AWGN channel, as if $\rvw(v)$ is AWGN conditionally independent of $\rvx$. Despite of its abstract form, such a general representation of the CL receiver architecture in fact covers many important use cases:

\begin{itemize}
    \item Non-Gaussian additive noise channels. The ``noise part'' $\rvw(v)$ is usually strictly independent of $\rvx$, but is non-Gaussian (see, e.g., \cite{lapidoth96it}). This case also includes channels with multiuser interference, to be treated in the case study in Sections \ref{sec:rate_analysis} and \ref{sec:coded_modulation}.
    \item Fading channels with pilot-assisted receiver (see, e.g., \cite{hassibi03it,tong04spm}). The received pilot plays as the side information $\rvv$, and an estimate of the fading state $\rvs$ based upon $\rvv$ is treated as the true fading state for coherent decoding. See also discussion in \cite[Sec. V-A]{wang22it} and \cite{pang21itw} for further elaboration.
    \item Channels whose transceivers have hardware imperfections and impairments (see, e.g., \cite{ochiai02com,zhang12com,bjornson14it}). Here the representation (\ref{eqn:channel-linearization}) is sometimes called ``Bussgang decomposition,'' a name stemming from a classical result \cite{bussgang52:rle,rowe82:bstj} due to Bussgang regarding the autocorrelation function of a continuous-time Gaussian signal undergoing memoryless nonlinear transformation.
\end{itemize}

The study of nearest neighbor decoding for general channels can be covered by the topic of mismatched decoding, i.e., decoding under some metric not matched to the actual channel transition law (i.e., conditional probability distributions); see \cite{lapidoth98tit,scarlett20now} and the brief outline in Section \ref{subsec:gnnd_review}. By modifying the Euclidean distance to incorporate general, possibly nonlinear, mappings of the channel output and the scaling coefficient of the channel input, we have proposed and studied generalized nearest neighbor decoding (GNND) in a recent work \cite{wang22it}. Intuitively, one may view GNND as a way of transforming the space where the original channel input and channel output reside, so that their ``distance'' after transformation becomes more adapted to the channel transition law. In \cite{wang22it} the GNND is optimized when the channel input is Gaussian, under the criterion of generalized mutual information (GMI), an extensively adopted performance measure in mismatched decoding; {for a concise introduction to mismatched decoding and GMI, see Section \ref{subsec:gnnd_review}.} An alternative perspective has been discussed in \cite{Kramer23arxiv}. The vector case of GNND aided by successive decoding has been studied in \cite{pang21itw}, thereby extending the renowned vertical Bell Laboratories layered space-time (V-BLAST) receiver \cite{wolniansky98ursi,tse05book} for multiple-input-multiple-output (MIMO) channels, wherein the linear minimum mean squared error (LMMSE)-successive interference cancellation (SIC) procedure turns out to be a special case of GNND when the receive side information of fading is perfect. A case of GNND for non-ergodic fading channels using outage probability as performance measure has been studied in \cite{shi22itw}.

In this paper, we study the GNND under general input constellations; {that is, any probability distribution of the input $\rvx$, instead of only Gaussian studied in \cite{wang22it}}. This is certainly relevant and important when one attempts to apply the GNND receiver architecture in practical systems, where the constellations are discrete. We derive criteria of the optimal GNND that maximizes the GMI, and the resulting optimality condition exhibits an interesting and insightful structural property of conditional moments matching; --- roughly speaking, an auxiliary probability distribution induced by the GNND should ``mimic'' the true a posteriori channel probability distribution in the sense that their first and second conditional moments match, respectively.

Based upon the established optimal GNND under general input constellations, we pick interference suppression in multiuser uplink as a use case, to study it for illustrating the benefit of GNND compared with CL. As discussed when enumerating possible use cases for the CL receiver architecture following (\ref{eqn:channel-linearization}), when considering an individual user, the non-Gaussian multiuser interference is the key challenge to handle. The LMMSE-SIC procedure as in the V-BLAST receiver architecture for MIMO channels still applies here \cite{tse05book,heathbook}, but it is optimal only when all the users employ Gaussian inputs. In many practical systems with a moderate number of users, due to complexity considerations, linear receivers such as matched filter (MF), zero-forcing (ZF) and LMMSE are used, without SIC. In such systems, when the system load (quantified by the ratio between the number of users and the number of receive antennas) gets heavy, say, in dense user-centric networks, cell-free networks, and other scenarios, linear receivers suffer from the excess interference induced by the large number of transmitted streams and usually perform poorly. Our evaluation of GMI under both the optimal GNND and the CL receivers indicates that, for discrete input constellations such as quadrature phase shift keying (QPSK), the CL approach incurs a noticeable rate loss compared with the channel mutual information, but the GNND approach is usually capable of achieving information rates nearly identical to the channel mutual information, even when the system load is heavy (i.e., when the number of users is comparable to or even larger than the number of receive antennas).

Beyond the GMI analysis, we further study the implementation of GNND for multiuser interference suppression using off-the-shelf codes. An attractive ``plug-and-play'' feature of GNND is that it can be implemented as a symbol-level memoryless preprocessing step, directly inserted between channel output and off-the-shelf decoder in legacy receivers, without any further modification of system. First, we conduct experiments for convolutional codes using Viterbi decoding, where the branch metric can be modified to genuinely realize GNND, CL, or ML decoding. The experimental results confirm that GNND exhibits evident gain compared with CL, and almost coincides with ML. Then, we further conduct experiments for a coded modulation scheme using low-density parity-check (LDPC) codes in standards. GNND is implemented via an additive noise channel representation so as to initialize the log-likelihood ratios (LLRs) fed into the belief propagation (BP) iterative procedure for LDPC decoding. To manage the computational cost of realizing the conditional expectation operator in GNND, a neural network based approximation implementation is employed. The experimental results again indicate substantial gain compared with CL, illustrating the potential of GNND as an improved receiver architecture.

The remaining part of this paper is organized as follows. Section \ref{sec:gnnd_general} briefly reviews background knowledge including mismatched decoding and GNND, and then establishes the optimality criteria for GNND under general input constellations. Section \ref{sec:rate_analysis} applies the derived results to study the case of multiuser interference suppression, revealing the information rate gain of GNND compared with CL. Section \ref{sec:coded_modulation} further studies the implementation of GNND, showcasing experimental results for convolutional codes using Viterbi decoding, and for coded modulation schemes using GNND-empowered LDPC codes. Finally, Section \ref{sec:conclusion} concludes this paper.

\section{Generalized Nearest Neighbor Decoding: Review and Optimal Solution under General Input Constellations}
\label{sec:gnnd_general}

In this section, we first provide a succinct review of existing results on GNND, summarizing \cite{wang22it}, and then derive the optimal GNND under general input constellations.

\subsection{Review of GNND}
\label{subsec:gnnd_review}

Mismatched decoding is a topic in information theory, dealing with the fundamental performance limit of channel transmission when the decoder is restricted to adopt specific decoding methods that are not necessarily capacity-achieving; for a detailed survey, see, e.g., \cite{lapidoth98tit,scarlett20now} and references therein. Consider a discrete-time memoryless channel with input $\rvx \in \mathcal{X}$ and output $\rvy \in \mathcal{Y}$, linked by a channel transition law $p_{\rvy|\rvx}(y|x)$, $\forall (x, y) \in \mathcal{X} \times \mathcal{Y}$. Consider a codebook $\mathcal{C} = \{\bm{x}(m): m = 1, \ldots, \lceil e^{NR}\rceil\}$ whose $m$-th codeword is $\bm{x}(m) = [x_1(m), x_2(m), \cdots, x_N(m)]$, $N$ denoting the code length and $R$ denoting the code rate (in nats/channel use). The ML decoder solves the following problem:
\begin{equation}
    \label{eqn:decoder-ML}
    \mathcal{D}_\mathrm{ML}: \hat{m} = \arg\max_{m \in \mathcal{M}} \sum_{n = 1}^N \ln p_{\rvy|\rvx}(y_n | x_n(m)),
\end{equation}
where $\mathcal{M} = \{1, \ldots, \lceil e^{NR}\rceil\}$ is the set of messages. Mismatched decoding treats the scenario where the decoder replaces $- \ln p_{\rvy|\rvx}(y|x)$ in (\ref{eqn:decoder-ML}) with some other function $d(x, y)$ called decoding metric, and the resulting mismatched decoder becomes:
\begin{equation}
    \label{eqn:decoder-d}
    \mathcal{D}_d: \hat{m} = \arg\min_{m \in \mathcal{M}} \sum_{n = 1}^N d(x_n(m), y_n).
\end{equation}
In (\ref{eqn:decoder-ML}) and (\ref{eqn:decoder-d}) we let ties be broken arbitrarily.

As discussed in the introduction, a decoding metric that has received much attention (see, e.g., \cite{lapidoth96it,lapidoth02it,weingarten04it,asyhari12it,zhang12com}) is the Euclidean distance $d(x, y) = |y - \alpha x|^2$ where $\alpha$ is a channel-dependent coefficient. This leads to a nearest neighbor decoder which seeks for the (scaled) codeword $\alpha \bm x(m)$, $m \in \mathcal{M}$, that has the smallest Euclidean distance to $\bm y$.

For general channels, the nearest neighbor decoder has still been widely adopted, leading to the well known CL receiver architecture. Moving one step further, GNND has been recently proposed in \cite{wang22it}, whose basic idea is to apply symbol-by-symbol memoryless transforms called processing function $g$ and scaling function $f$, so as to modify the decoding metric into
\begin{equation}
\label{eqn:GNND}
    d(x, y) = |g(y) - f(y) x|^2.
\end{equation}
\begin{rem}
    \label{rem:augmented-output}
    As described in the introduction, the channel model in \cite{wang22it} can further include a state $\rvs \in \mathcal{S}$ and a receive side information $\rvv \in \mathcal{V}$ satisfying the Markov chain $\rvv \leftrightarrow \rvs \leftrightarrow (\rvx, \rvy)$. To handle such a situation, in (\ref{eqn:GNND}) we may simply augment the variable $y$ to $(y, v)$.
\end{rem}

For a given decoding metric $d$, we call a code rate $R$ achievable, if there exists a sequence of codebooks such that the maximal decoding error probability for a decoder adopting the mismatched decoding rule (\ref{eqn:decoder-d}) asymptotically vanishes as $N$ grows without bound. We define the mismatch capacity as the supremum of achievable code rates with respect to $d$.

Only when the decoding metric is matched to the channel, i.e., $d(x, y) \propto -\ln p_{\rvy|\rvx}(y|x)$, $\forall (x, y) \in \mathcal{X}\times \mathcal{Y}$, the channel mutual information $I(\rvx; \rvy)$ is achievable with a given input probability distribution $p_\rvx$ and the channel capacity is further obtained by optimizing $p_\rvx$. In general, when $d$ is mismatched to the channel, the mismatch capacity has remained an open problem to date \cite{scarlett20now}.

Several lower bounds of the mismatch capacity under given input probability distribution are available, including the GMI \cite{stiglitz66,kaplan93aeu}, the LM rate \cite{csiszar81,hui83}, their multi-letter improvements \cite{csiszar95}, and rates based on multiuser coding techniques \cite{lapidoth96,somekh-baruch15,scarlett16}. Among these lower bounds, the GMI is the simplest and by far the most extensively used in applications, based on codebook ensembles consisting of independent and identically distributed (i.i.d.) random variables. {The GMI is a threshold rate for i.i.d. codebook ensembles, in the sense that the ensemble average decoding error probability vanishes asymptotically below it, and tends to one asymptotically above it, as the code length grows without bound (see, e.g., \cite[Sec. III]{lapidoth02it}).}

We adopt the GMI as the performance measure, and evaluate it using the following general expression \cite[Eqn. (12)]{ganti00it}:
\begin{align}
\label{eqn:dual form}
    &I_\mathrm{GMI} = \sup_{\theta < 0} \sum_{(x, y) \in \mathcal{X} \times \mathcal{Y}} p_\rvx(x) p_{\rvy|\rvx}(y|x) \nonumber\\
    & \quad\quad\quad\quad\quad\quad \times \log \frac{e^{\theta d(x, y)}}{\sum_{x' \in \mathcal{X}}p_\rvx(x') e^{\theta d(x', y)}},
\end{align}
under the i.i.d. codebook ensemble with probability distribution $p_\rvx$. Note that (\ref{eqn:dual form}) also applies to channels with continuous input/output alphabets, with summation replaced by integration accordingly. Furthermore, since we consider GNND (\ref{eqn:GNND}), we seek to maximize the GMI in (\ref{eqn:dual form}) by optimizing the processing function $g$ and the scaling function $f$.

When $p_\rvx$ is chosen to be circularly symmetric complex Gaussian with average power $P$, i.e., $\rvx \sim \mathcal{CN}(0, P)$, and when the channel output is augmented to $(\rvy, \rvv)$ (see Remark \ref{rem:augmented-output}), it has been shown in \cite[Thm. 1]{wang22it} that the maximized GMI of GNND is given by
\begin{equation}
    \label{eqn:GMI-opt}
    I_{\mathrm{GMI}, \mathrm{opt}} = \mathbf{E} \left[ \log \frac{P}{\omega (\rvy, \rvv)} \right],
\end{equation}
where $\omega(y, v)$ is the conditional variance of $\rvx$ with respect to $p_{\rvx|\rvy, \rvv}(x|y, v)$, i.e.,
\begin{equation}
    \omega (y, v) = \mathbf{var}\left[\rvx | y, v\right] := \mathbf{E} \left[|\rvx|^2 | y, v\right] - \left|\mathbf{E}\left[\rvx | y, v\right]\right|^2,
\end{equation}
and the corresponding optimal solution of $(g, f)$ is
\begin{align}
    \label{eqn:decoder-GNND-optimal-g}
    g(y, v) &= \frac{1}{\sqrt{(P - \omega(y, v)) \omega(y, v)}} \mathbf{E}[\rvx|y, v],\\
    \label{eqn:decoder-GNND-optimal-f}
    f(y, v) &= \frac{\sqrt{P - \omega(y, v)}}{P \sqrt{\omega(y, v)}}.
\end{align}
We note that multiplying both $g(y, v)$ and $f(y, v)$ in (\ref{eqn:decoder-GNND-optimal-g}) and (\ref{eqn:decoder-GNND-optimal-f}) with a common constant simultaneously does not affect their optimality.

On the other hand, it has been shown in \cite[Prop. 4]{wang22it} that the CL representation (\ref{eqn:channel-linearization}) achieves the following GMI under $\rvx \sim \mathcal{CN}(0, P)$:
\begin{equation}
    \label{eqn:GMI-lin}
    I_{\mathrm{GMI}, \mathrm{lin}} = \mathbf{E}\left[\log \frac{P}{\mathrm{lmmse}_\rvv}\right],
\end{equation}
where $\mathrm{lmmse}_\rvv := P - \mathbf{E}[\rvx^\ast \rvy | \rvv]^\ast \mathbf{E}[\rvy \rvy^\ast | \rvv]^{-1} \mathbf{E}[\rvx^\ast \rvy | \rvv]$
is the conditional MMSE of the LMMSE estimator of $\rvx$ upon observing $\rvy$, conditioned upon $\rvv$.

In general $I_{\mathrm{GMI}, \mathrm{opt}} \geq I_{\mathrm{GMI}, \mathrm{lin}}$ holds, and their gap characterizes the information rate gain from optimizing GNND. {For several representative cases, including fading channels with imperfect receive side information and channels with output quantization, the gap can be substantial; see \cite[Sec. V]{wang22it} \cite{pang21itw}.}

\subsection{GNND under General Input Constellations}
\label{subsec:gnnd_general}

Having briefly reviewed the background of mismatched decoding and GNND in the previous subsection, here we proceed to study the optimal GNND under general input constellations.

In order to keep the notation consistent with the GNND in the previous subsection, we consider channels with state $\rvs$ and receive side information $\rvv$, so that the augmented channel output is $(\rvy, \rvv)$, as discussed in Remark \ref{rem:augmented-output}. From the general expression of the GMI (\ref{eqn:dual form}), for given $g$ and $f$, the GMI with GNND is given by
\begin{align}
    \label{eqn:gmi-general-constellation}
    &I_{\mathrm{GMI},g,f} = \sup_{\theta < 0} \Bigg\{ \theta \sum_{(x, y, v) \in \mathcal{X}\times \mathcal{Y}\times \mathcal{V}} p_\rvx(x) p_{\rvy, \rvv|\rvx}(y, v|x) \nonumber\\
    & \times |f(y, v) x - g(y, v)|^2 - \sum_{(y, v) \in \mathcal{Y} \times \mathcal{V}} p_{\rvy, \rvv}(y, v) \nonumber\\
    & \times \log \sum_{x' \in \mathcal{X}} p_\rvx(x') e^{\theta |f(y, v) x' - g(y, v)|^2}\Bigg\}.
\end{align}

Our following result provides a general characterization of the optimal $(g, f)$.

\begin{thm}
\label{thm:GMI-general-input}
For a channel transmission model with scalar input $\rvx \in \mathcal{X} \subseteq \mathbb{C}$, state $\rvs \in \mathcal{S}$, receive side information $\rvv \in \mathcal{V}$ and output $\rvy \in \mathcal{Y}$, under input probability distribution $p_\rvx$,
the optimal $(g, f)$ that maximizes the GMI of GNND should satisfy, for each $(y, v) \in \mathcal{Y} \times \mathcal{V}$:
\begin{align}
\label{eqn:gmi-maximizing-1st}
\frac{\mathbf{E} \left[\rvx e^{-|f(y, v)\rvx - g(y, v)|^2}\right]}{\mathbf{E} \left[ e^{-|f(y, v)\rvx - g(y, v)|^2 }\right]} &= \mathbf{E}[\rvx| y, v],\\
\label{eqn:gmi-maximizing-2nd}
\frac{\mathbf{E} \left[|\rvx|^2 e^{-|f(y, v)\rvx - g(y, v)|^2}\right]}{\mathbf{E} \left[ e^{-|f(y, v)\rvx - g(y, v)|^2 }\right]} &= \mathbf{E}[|\rvx|^2|y, v],
\end{align}
{where the expectation on the left hand side is taken with respect to the probability distribution of $\rvx$.}
\end{thm}
\textit{Proof}: See Appendix \ref{sec:proof of theorem 1}. $\Box$

\begin{rem}
    \label{rem:matching-condition}
    For each $(y, v) \in \mathcal{Y} \times \mathcal{V}$, conditions (\ref{eqn:gmi-maximizing-1st}) and (\ref{eqn:gmi-maximizing-2nd}) specify a pair of numbers $(g(y, v), f(y, v))$. Over $\mathcal{Y} \times \mathcal{V}$, $\{(g(y, v), f(y, v)): (y, v) \in \mathcal{Y} \times \mathcal{V}\}$ thus specify a pair of mappings $(g, f)$. By defining an auxiliary probability distribution as
    \begin{equation}
        \label{eqn:auxiliary-p}
        \tilde{p}_{\rvx|\rvy, \rvv}(x|y, v) = \frac{e^{-|f(y, v) x - g(y, v)|^2} p_\rvx(x) }{\mathbf{E} \left[e^{-|f(y, v)\rvx - g(y, v)|^2}\right]},
    \end{equation}
    we can rewrite conditions (\ref{eqn:gmi-maximizing-1st}) and (\ref{eqn:gmi-maximizing-2nd}) in the following form of conditional moments matching:
    \begin{align}
        \label{eqn:moment-matching-1}
        \mathbf{E}_{\tilde{p}_{\rvx|\rvy, \rvv}}[\rvx|y, v] &= \mathbf{E}_{p_{\rvx|\rvy, \rvv}}[\rvx|y, v],\\
        \label{eqn:moment-matching-2}
        \mathbf{E}_{\tilde{p}_{\rvx |\rvy, \rvv}}[|\rvx|^2|y, v] &= \mathbf{E}_{p_{\rvx |\rvy, \rvv}}[|\rvx|^2|y, v].
    \end{align}
    This interpretation provides a crucial insight into the behavior of the optimal GNND; that is, the optimal $(g, f)$ should render the auxiliary probability distribution $\tilde{p}_{\rvx|\rvy, \rvv}$ in (\ref{eqn:auxiliary-p}) to ``mimic'' the true a posteriori channel probability distribution $p_{\rvx|\rvy, \rvv}$ in the sense that their first and second conditional moments match, respectively.
\end{rem}

\begin{rem}
    \label{rem:kramer-anticipation-false}
    From conditions (\ref{eqn:gmi-maximizing-1st}) and (\ref{eqn:gmi-maximizing-2nd}), it is clear that the optimal $(g, f)$ depends upon the conditional expectations $\mathbf{E}[\rvx|y, v]$ and $\mathbf{E}[|\rvx|^2|y, v]$ only, which are estimation-theoretic, rather than decision-theoretic, quantities. This is consistent with our earlier results for Gaussian input, i.e., (\ref{eqn:decoder-GNND-optimal-g}) and (\ref{eqn:decoder-GNND-optimal-f}).
\end{rem}

\begin{rem}
    \label{rem:gaussian-revisit}
    From the conditional moments matching criteria (\ref{eqn:moment-matching-1}) and (\ref{eqn:moment-matching-2}), we can revisit the optimal $(g, f)$ in (\ref{eqn:decoder-GNND-optimal-g}) and (\ref{eqn:decoder-GNND-optimal-f}) under Gaussian input. Applying $\rvx \sim \mathcal{CN}(0, P)$ to (\ref{eqn:auxiliary-p}), and conducting some algebraic manipulations, we obtain
        \begin{eqnarray}
    	\tilde{p}_{\rvx|\rvy, \rvv}(x|y, v) = \frac{1}{\pi\nu(y, v)} e^{- |x - \nu(y, v) f^\dag (y, v) g(y, v)|^2/\nu(y, v)},
    \end{eqnarray}
    where $1/\nu(y, v) = |f(y, v)|^2 + 1/P$. Therefore, (\ref{eqn:moment-matching-1}) and (\ref{eqn:moment-matching-2}) become
    \begin{align}
        \mathbf{E}_{\tilde{p}_{\rvx|\rvy, \rvv}}[\rvx|y, v] &= \nu(y, v) f^\dag (y, v) g(y, v) \nonumber\\
        &= \mathbf{E}_{p_{\rvx|\rvy, \rvv}}[\rvx|y, v],\\
        \mathbf{E}_{\tilde{p}_{\rvx |\rvy, \rvv}}[|\rvx|^2|y, v] &= \nu(y, v) + |\mathbf{E}_{\tilde{p}_{\rvx|\rvy, \rvv}}[\rvx|y, v]|^2 \nonumber\\
        &= \nu(y, v) + |\mathbf{E}_{p_{\rvx|\rvy, \rvv}}[\rvx|y, v]|^2 \nonumber\\
        &= \mathbf{E}_{p_{\rvx |\rvy, \rvv}}[|\rvx|^2|y, v],
    \end{align}
    respectively; that is,
    \begin{align}
        \frac{f^\dag (y, v) g(y, v)}{|f(y, v)|^2 + 1/P} &= \mathbf{E}[\rvx|y, v],\\
        \frac{1}{|f(y, v)|^2 + 1/P} &= \mathbf{E}[|\rvx|^2|y, v] - |\mathbf{E}[\rvx|y, v]|^2.
    \end{align}
    These immediately lead to one choice of $(g,f)$ pair
    \begin{align}
        g(y, v) &= \sqrt{\frac{P}{(P - \omega(y, v))\omega(y, v)}} \mathbf{E}[\rvx|y, v],\\
        f(y, v) &= \sqrt{\frac{P - \omega(y, v)}{P \omega(y, v)}},
    \end{align}
    which are equivalent to (\ref{eqn:decoder-GNND-optimal-g}) and (\ref{eqn:decoder-GNND-optimal-f}), with scaling.
\end{rem}

The following result characterizes the gap between the channel mutual information and the GMI under the optimal $(g, f)$ in Theorem \ref{thm:GMI-general-input}.

\begin{cor}
    \label{cor:gap-mi-gmi}
    For the optimal $(g, f)$ that maximizes the GMI as characterized by Theorem \ref{thm:GMI-general-input}, the gap between the channel mutual information and the resulting GMI is given by
    \begin{equation}
        I(\rvx; \rvy, \rvv) - I_\mathrm{GMI} = D(p_{\rvx|\rvy, \rvv} \| \tilde{p}_{\rvx|\rvy, \rvv} | p_{\rvy, \rvv}),
    \end{equation}
    where $\tilde{p}_{\rvx|\rvy, \rvv}$ is given by (\ref{eqn:auxiliary-p}) in Remark \ref{rem:matching-condition}.
\end{cor}
\textit{Proof}: See Appendix \ref{sec:proof of corollary 1}. $\Box$

\begin{rem}
    \label{rem:gap-mi-gmi}
    We see that the loss due to using GNND is the conditional Kullback-Leibler divergence between the true channel posteriori probability distribution $p_{\rvx|\rvy, \rvv}$ and the auxiliary probability distribution $\tilde{p}_{\rvx|\rvy, \rvv}$, averaged over $p_{\rvy, \rvv}$. In light of the conditional moments matching criteria (\ref{eqn:moment-matching-1}) and (\ref{eqn:moment-matching-2}), it is expected that this gap $D(p_{\rvx|\rvy, \rvv} \| \tilde{p}_{\rvx|\rvy, \rvv} | p_{\rvy, \rvv})$ is often relatively small, as will be illustrated by our case study in Section \ref{sec:rate_analysis}.
\end{rem}

We proceed to investigate the specific case where $p_\rvx$ is QPSK. Under an average power constraint $P$, the equiprobable constellation points of QPSK are
\begin{align*}
    & \{a_1, a_2, a_3, a_4\} = \Big\{\sqrt{P/2} (1 + \jmath), \\
    & \sqrt{P/2} (1 - \jmath), \sqrt{P/2} (-1 + \jmath), \sqrt{P/2} (-1 - \jmath)\Big\},  
\end{align*}
respectively. Applying Theorem \ref{thm:GMI-general-input}, the maximized GMI and its associated decoding metric of GNND are given by the following result.

\begin{thm}
    \label{thm:GMI-opt-qpsk}
    For QPSK input under an average power constraint $P$, the maximized GMI of GNND is given by (\ref{eqn:GMI-opt-qpsk}),
        \begin{figure*}
    \begin{align}
        \label{eqn:GMI-opt-qpsk}
        I_{\mathrm{GMI}} &= \mathbf{E} \bigg[ \sqrt{\frac{2}{P}} \Re\{\mathbf{\mathbf{E}}[\rvx| \rvy, \rvv]\} \mathrm{artanh}\bigg(\sqrt{\frac{2}{P}} \Re\{\mathbf{\mathbf{E}}[\rvx| \rvy, \rvv]\}\bigg) + \sqrt{\frac{2}{P}} \Im\{\mathbf{\mathbf{E}}[\rvx| \rvy, \rvv]\} \mathrm{artanh}\bigg(\sqrt{\frac{2}{P}} \Im\{\mathbf{\mathbf{E}}[\rvx| \rvy, \rvv]\} \bigg)\nonumber\\
        & \quad\quad - \log \bigg( \cosh\bigg( \mathrm{artanh}\bigg(  \sqrt{\frac{2}{P}} \Re\{\mathbf{\mathbf{E}}[\rvx| \rvy, \rvv]\} \bigg)\bigg)\cosh\bigg(\mathrm{artanh} \bigg(\sqrt{\frac{2}{P}} \Im\{\mathbf{\mathbf{E}}[\rvx| \rvy, \rvv]\}\bigg)\bigg) \bigg) \bigg],
    \end{align}
    \end{figure*}
    and the associated decoding metric is
    \begin{align}
        \label{eqn:GNND-qpsk-opt}
        d(x, (y, v)) &= \bigg| \dfrac{1}{\sqrt{2P}} \mathrm{artanh} \sqrt{\frac{2}{P}} \Re\{\mathbf{\mathbf{E}}[\rvx| y, v]\} + \nonumber\\
          &\quad\ \  \dfrac{\jmath}{\sqrt{2P}} \mathrm{artanh} \sqrt{\frac{2}{P}} \Im\{\mathbf{\mathbf{E}}[\rvx| y, v]\} - x\bigg|^2.
    \end{align}
\end{thm}
\textit{Proof}: See Appendix \ref{sec:proof of theorem 2}. $\Box$

Inspecting the decoding metric \eqref{eqn:GNND-qpsk-opt}, we may view the processed channel output $\frac{1}{\sqrt{2P}} \mathrm{artanh} \sqrt{\frac{2}{P}} \Re\{\mathbf{\mathbf{E}}[\rvx| y, v]\} + \frac{\jmath}{\sqrt{2P}} \mathrm{artanh} \sqrt{\frac{2}{P}} \Im\{\mathbf{\mathbf{E}}[\rvx| y, v]\}$ as an estimate of $x$. Since the purpose of this estimate is decoding, we may call it the ``GNND estimate''; see Figures \ref{fig:x-decomposition} and \ref{fig:x-decomposition-learned} in Section \ref{sec:coded_modulation} for some examples illustrating its advantage over the LMMSE estimate used in CL.

\subsection{GMI with CL}
\label{subsec:GNND-channel-linearization}

For comparison, we briefly review the results on GMI with CL, based on the channel representation (\ref{eqn:channel-linearization}). Note that the validity of the key property of $\rvw(v)$ and $\rvx$ being conditionally uncorrelated does not rely upon the probability distribution of $\rvx$. Therefore the analysis in \cite[Sec. IV-C]{wang22it} still applies here. For the channel representation (\ref{eqn:channel-linearization}),
\begin{equation}
    \rvy = \frac{\mathbf{E}\left[\rvx^\dag \rvy | v\right]}{P} \rvx + \rvw(v), 
\end{equation}
the covariance matrix of $\rvw(v)$ is $\Delta(v) = \mathbf{E}[\rvy \rvy^\dag|v] - \frac{1}{P} \mathbf{E}[\rvx^\dag \rvy|v] \mathbf{E}[\rvx^\dag \rvy|v]^\dag$. The idea of CL is to postulate $\rvw(v)$ as Gaussian conditionally independent of $\rvx$ given $v$. Correspondingly the optimal receiver first removes the spatial correlation in $\rvw(v)$ by a whitening matrix $\Delta(v)^{-1/2}$, and then applies a maximum ratio combiner, yielding the channel model
\begin{align}
    &\frac{\mathbf{E}[\rvx^\dag \rvy|v]^\dag \Delta(v)^{-1}}{\sqrt{\mathbf{E}[\rvx^\dag \rvy|v]^\dag \Delta(v)^{-1} \mathbf{E}[\rvx^\dag \rvy|v]}} \rvy = \nonumber\\
    &\quad\quad\frac{\mathbf{E}[\rvx^\dag \rvy|v]^\dag \Delta(v)^{-1} \mathbf{E}[\rvx^\dag \rvy|v]}{P\sqrt{\mathbf{E}[\rvx^\dag \rvy|v]^\dag \Delta(v)^{-1} \mathbf{E}[\rvx^\dag \rvy|v]}} \rvx + \tilde{\rvw},
\end{align}
where $\tilde{\rvw}$ is postulated as $\mathcal{CN}(0, 1)$. The resulting decoding metric is thus
\begin{align}
\label{eqn:decoding-metric-CL}
    &d(x, (y, v)) = \frac{1}{{\mathbf{E}[\rvx^\dag \rvy|v]^\dag \Delta(v)^{-1} \mathbf{E}[\rvx^\dag \rvy|v]}} \times \nonumber\\
    &\Big| \mathbf{E}[\rvx^\dag \rvy|v]^\dag \Delta(v)^{-1} \left(y - \mathbf{E}[\rvx^\dag \rvy|v] {x}/{P}\right) \Big|^2,
\end{align}
a weighted nearest neighbor rule, and can be shown via an exercise of the matrix inversion lemma to be equivalent to the form given by \cite[Prop. 4]{wang22it}.

One can then evaluate the GMI with CL using (\ref{eqn:dual form}) and (\ref{eqn:decoding-metric-CL}). As an example, for the special case of QPSK input without receive side information, the resulting expression is given by (\ref{eqn:gmi-qpsk-lin}), which will be used in our case study in the following sections.
\begin{figure*}
\begin{align}
    \label{eqn:gmi-qpsk-lin}
    I_{\mathrm{GMI}, \mathrm{lin}} &= \sup_{\theta<0} \Bigg\{  -2\theta \Re\left\{\mathbf{E}\left[\rvy^\dag \mathbf E\left[	{\rvw} {\rvw}^\dag \right]^{-1} \dfrac{\mathbf E[\rvx^\dag \rvy]}{P} \rvx \right]\right\} - \mathbf{E} \Bigg[\log \Bigg\{ \cosh\sqrt{2P} \theta \Re\left\{\rvy^\dag \mathbf E\left[{\rvw} {\rvw}^\dag \right]^{-1} \dfrac{\mathbf E[\rvx^\dag \rvy]}{P}\right\}\nonumber\\
    & \times \cosh\sqrt{2P} \theta \Im\left\{\rvy^\dag \mathbf E\left[{\rvw}	{\rvw}^\dag \right]^{-1} \dfrac{\mathbf E[\rvx^\dag \rvy]}{P}\right\}\Bigg\}\Bigg]\Bigg\}.
\end{align}
\end{figure*}

\section{GNND for Multiuser Interference Suppression}
\label{sec:rate_analysis}

In this section, we conduct a case study of interference suppression, to illustrate the potential of GNND. Consider a multiuser channel with $K$ non-cooperating single-antenna transmitters and a centralized receiver with $L$ receive antennas,
\begin{equation}
\label{eqn:gaussian-fading-channel}
    \rvy = \sum_{k = 1}^K {\bm h}_k \rvx_k + \rvz,
\end{equation}
where $\bm h_k \in \mathbb{C}^{L \times 1}$ denotes the channel gain coefficients between user $k$ and the $L$ receive antennas, and $\rvz \sim \mathcal{CN}(\mathbf{0}, \sigma^2 \mathbf{I}_{L\times L})$ denotes the AWGN at the receiver. The $K$ channel inputs $\rvx = [\rvx_1, \ldots, \rvx_K]$ are mutually independent and drawn from an input constellation, with per-transmitter average power constraint $\mathbf{E} [|\rvx_k|^2] = P_k$, $k = 1, \ldots, K$. We assume that the channel gain coefficients $[\bm h_1, \ldots, \bm h_K]$ are fixed throughout the transmission and are known at the decoder. Therefore they should not be treated as the receive side information. In the numerical experiments in Section \ref{sec:coded_modulation}, we will also study the effect of imperfect knowledge of $[\bm h_1, \ldots, \bm h_K]$ on the performance of GNND.

We consider two types of receiver, without and with SIC.

\begin{itemize}
	\item Receiver without SIC: The decoder conducts single-user decoding for each user's message separately in parallel, viewing other users' codewords as interference. Considering $\rvx_k$, its GMI with GNND and CL can then be evaluated according to the analysis in Section \ref{sec:gnnd_general}, with $\rvx$ replaced by $\rvx_k$.
	\item Receiver with SIC: The decoder decodes users' messages sequentially, treating previously decoded users' codewords as receive side information; that is, when decoding $\rvx_k$, the decoder augments its channel output from $\rvy$ to $(\rvy, \rvx_1, \ldots, \rvx_{k - 1})$, and then conducts GNND or CL, respectively. This is a generalization of the well-known V-BLAST receiver architecture: when all users' inputs are Gaussian, GNND and CL are equivalent and both exactly correspond to the LMMSE step in V-BLAST \cite{pang21itw}; but when users adopt general input constellations, GNND turns out to be more effective than CL in handling the multiuser interference, as will be shown in the sequel.
\end{itemize}

The GMI of user $k$ is denoted as $I_{\mathrm{GMI}, k}$, and the sum GMI is $I_\mathrm{GMI} = \sum_{k = 1}^K I_{\mathrm{GMI}, k}$.

We study numerical results on the GMI with GNND and CL, for receivers without and with SIC. We let all users adopt QPSK with equal power $P_k = P/K$ ($k = 1, \ldots, K$) and measure the system signal-to-noise ratio (SNR) as $P/\sigma^2$. For this setup, it can be shown that if $K = 1$ (i.e., the single-user special case), then we have $I(\rvx; \rvy) = I_{\mathrm{GMI}} = I_{\mathrm{GMI}, \mathrm{lin}}$. We hence consider various $(K, L)$ pairs with $K \geq 2$. The channel gain coefficients follow i.i.d. zero-mean unit-variance circularly symmetric complex Gaussian distribution and stay fixed throughout the transmission of a codeword, i.e., a quasi-static Rayleigh fading model.

\begin{figure*}[ht]
	\centering
		\subfigure[$K=4$, $L=2$]{
		\includegraphics[width=0.31\textwidth]
		{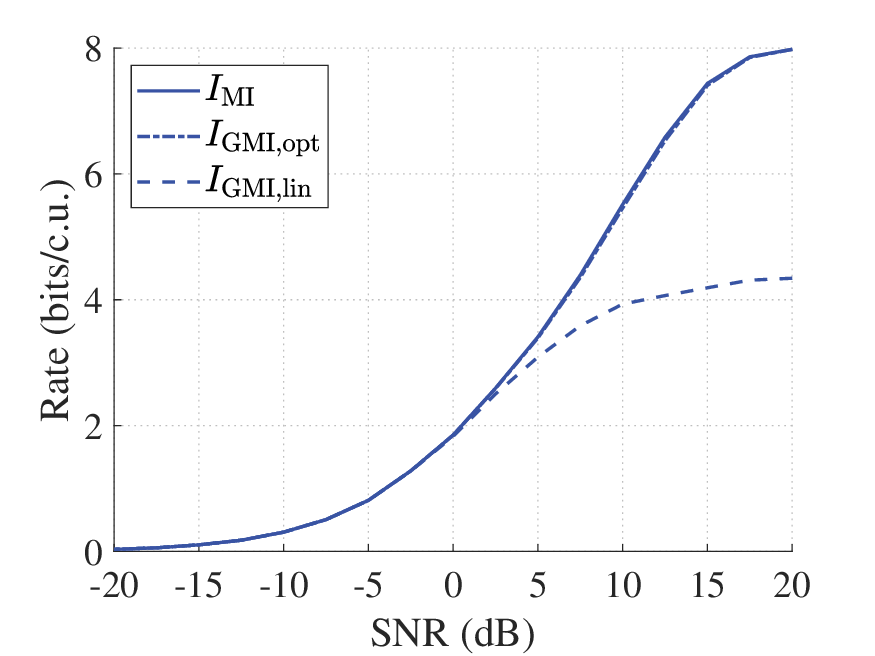}}
	\subfigure[$K=4$, $L=4$]{
		\includegraphics[width=0.31\textwidth]
		{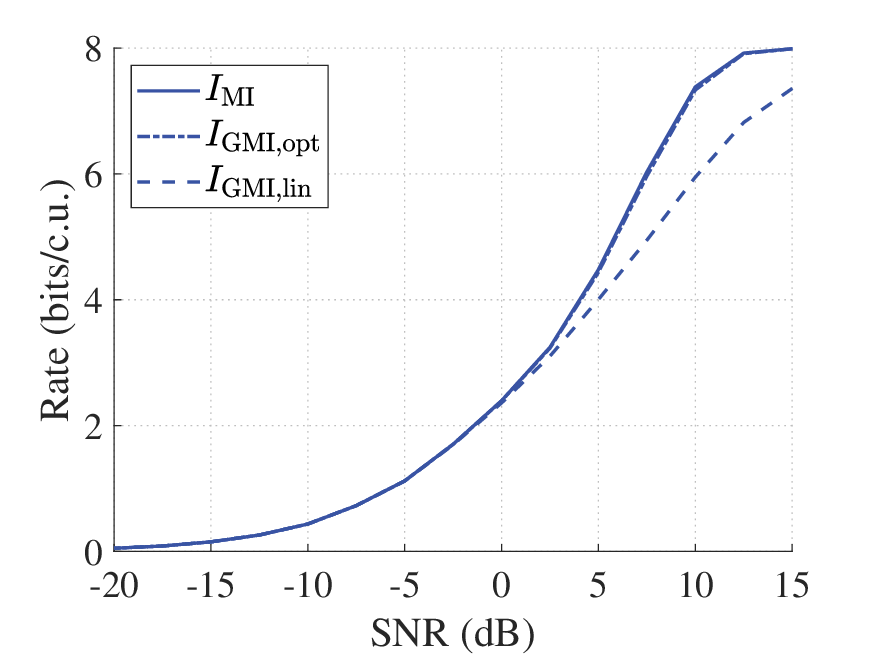}}
	\subfigure[$K=4$, $L=8$]{
		\includegraphics[width=0.31\textwidth]
		{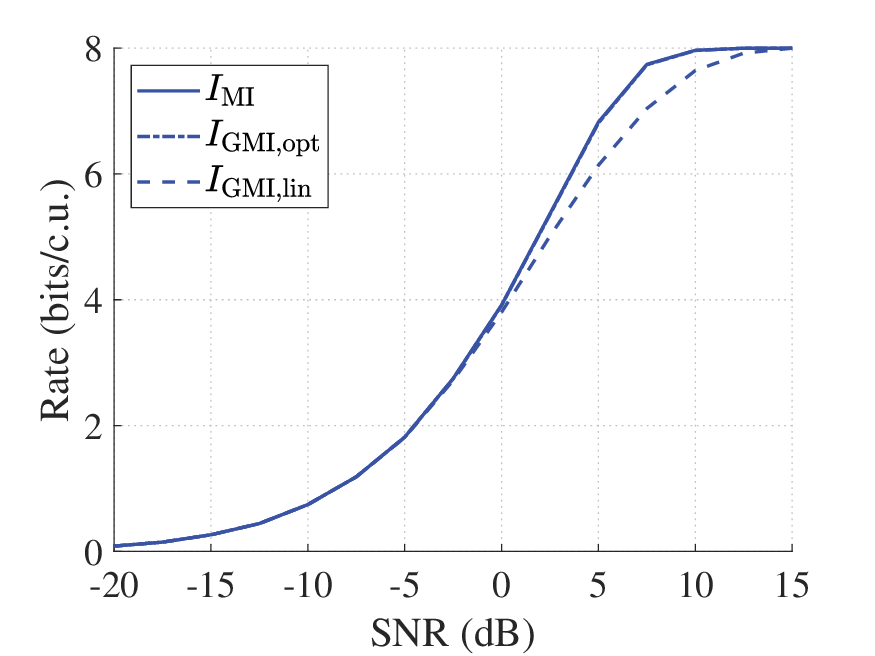}}
	\subfigure[$K=8$, $L=4$]{
		\includegraphics[width=0.31\textwidth]
{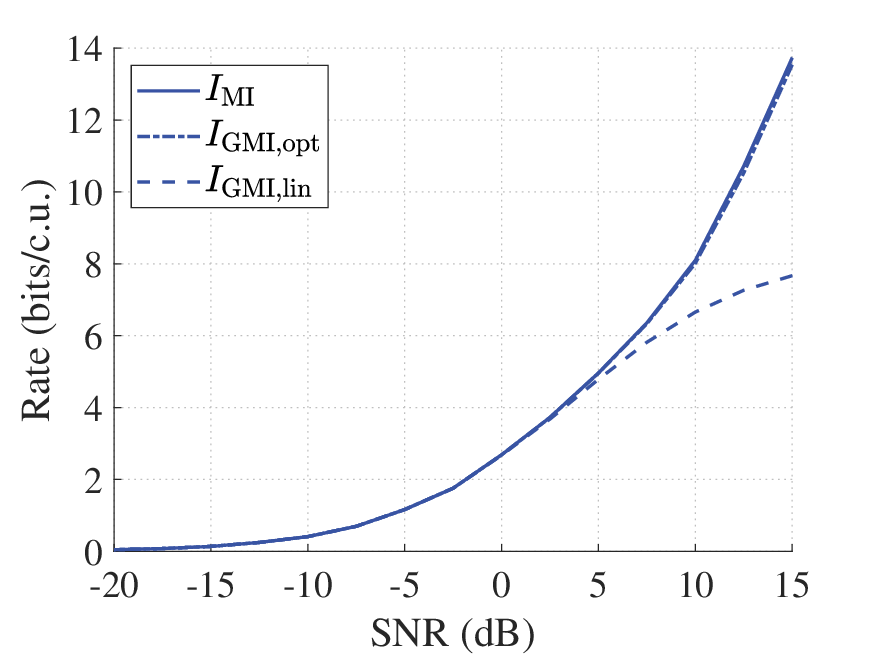}}
	\subfigure[$K=8$, $L=8$]{
		\includegraphics[width=0.31\textwidth]
		{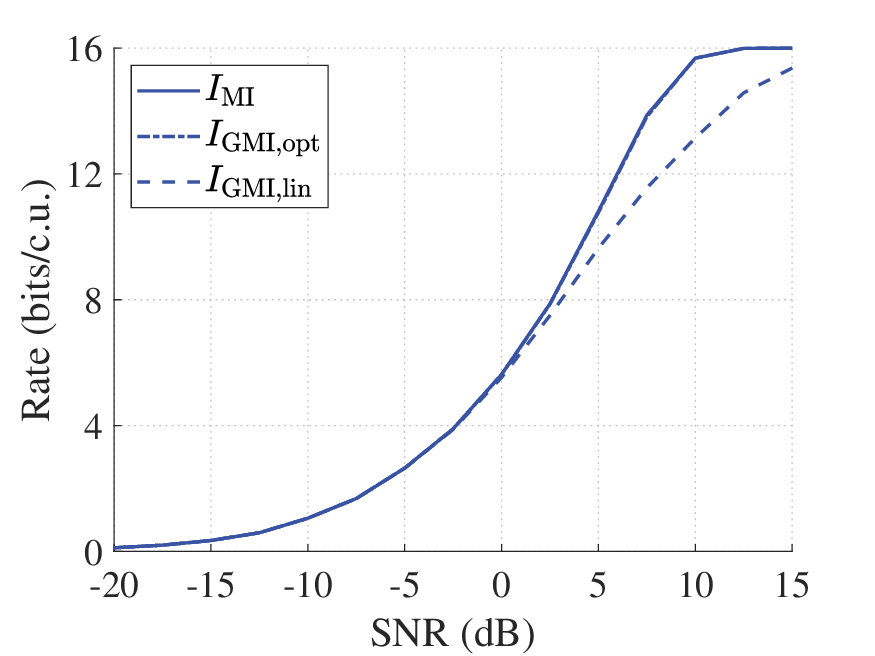}}
	\subfigure[$K=8$, $L=16$]{
		\includegraphics[width=0.31\textwidth]
		{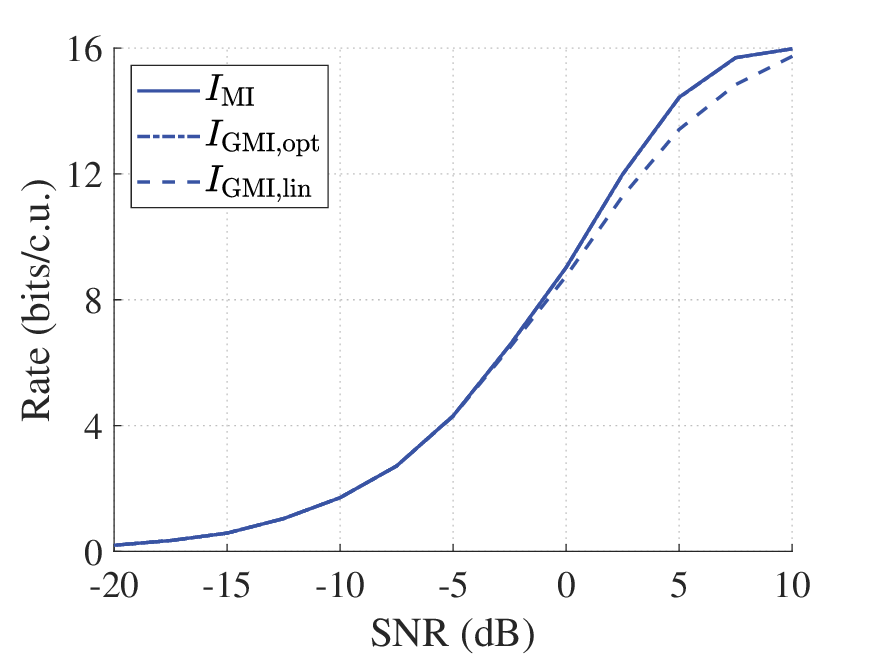}}
  	\caption{Sum GMIs and mutual information for multiuser channel \eqref{eqn:gaussian-fading-channel} under GNND without SIC.}
	\label{fig:gmi-perfect-MU-MIOM}
\end{figure*}

For receiver without SIC, we plot in Figure \ref{fig:gmi-perfect-MU-MIOM} the sum GMIs with GNND and CL, respectively, as well as the sum single-user mutual information $I_\mathrm{MI} = \sum_{k = 1}^K I(\rvx_k; \rvy)$, for comparison. We observe that, the GMI with GNND and the mutual information are surprisingly close throughout the range of SNR considered. In fact, their gap does not exceed $0.1$ bits/c.u. at SNR $10$ dB, nearly invisible in the figure. This phenomenon is inline with the analysis in Corollary \ref{cor:gap-mi-gmi}, suggesting that the auxiliary probability distribution $\tilde{p}_{\rvx_k|\rvy}$ is close to the true a posteriori channel probability distribution $p_{\rvx_k|\rvy}$ in terms of conditional Kullback-Leibler distance. On the other hand, the gap between the GMI with GNND and that with CL is clearly noticeable for the range of SNR of the most practical interest, and is the most significant for overloaded systems (e.g., $(K, L) = (4, 2)$ or $(8, 4)$). That the CL receiver fails to work for overloaded systems is no surprise because of the limitation of linear processing for under-determined systems, but that the GNND receiver still performs almost without loss is interesting and reassuring.

\begin{figure*}[ht]
	\centering
	\subfigure[$K=4$, $L=2$]{
	\includegraphics[width=0.31\textwidth]
	{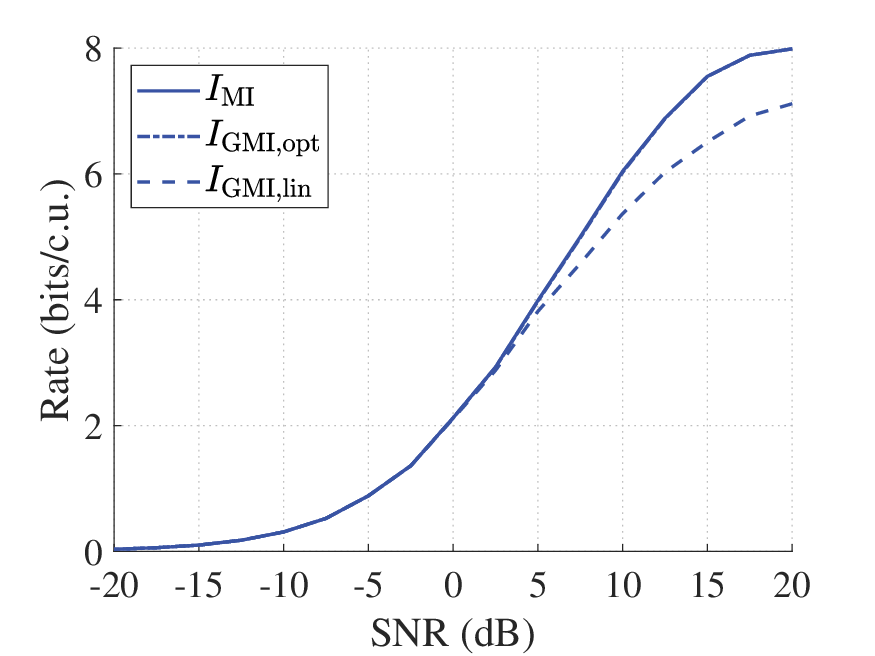}}
\subfigure[$K=4$, $L=4$]{
	\includegraphics[width=0.31\textwidth]
	{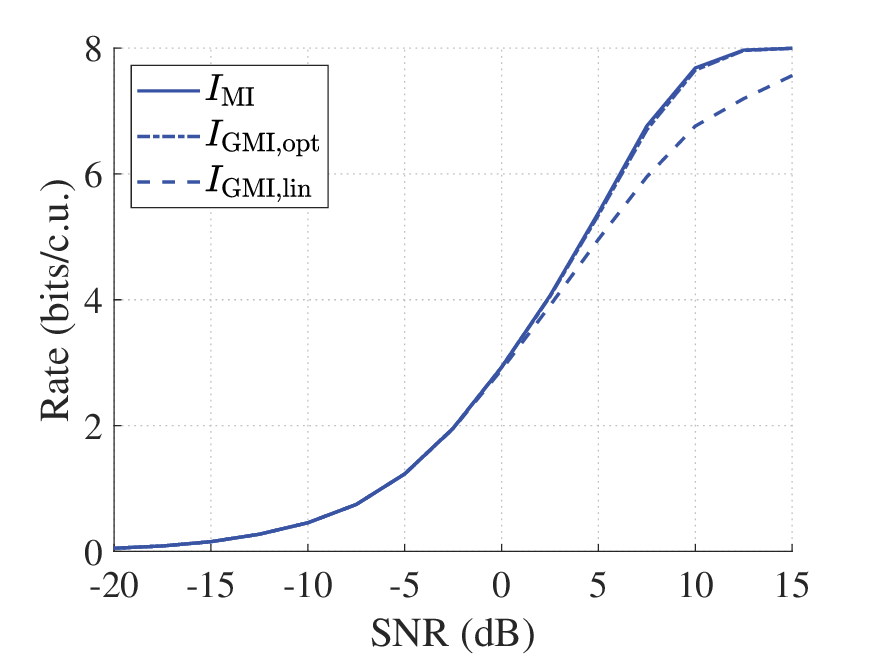}}
\subfigure[$K=4$, $L=8$]{
	\includegraphics[width=0.31\textwidth]
	{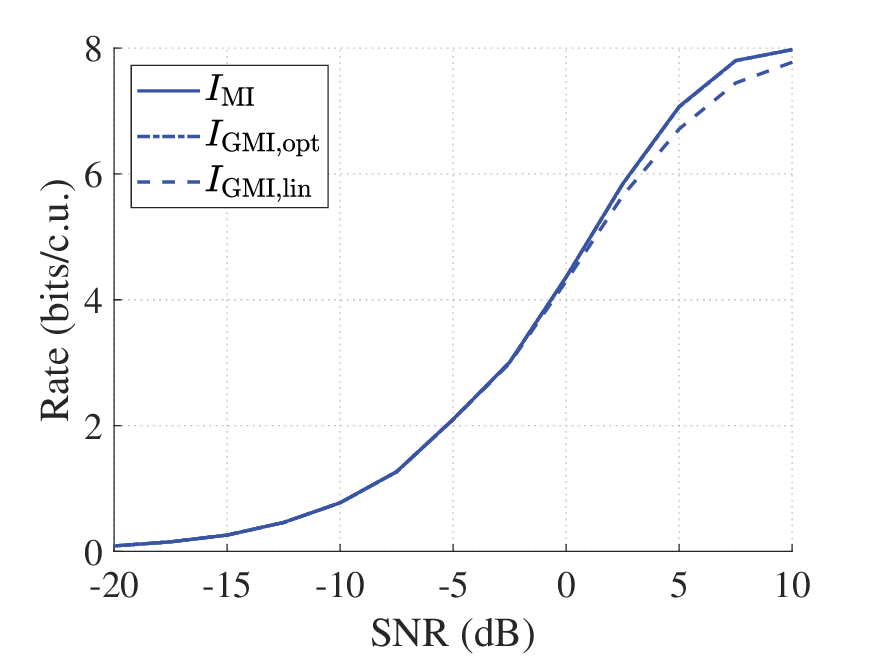}}
\subfigure[$K=8$, $L=4$]{
	\includegraphics[width=0.31\textwidth]
	{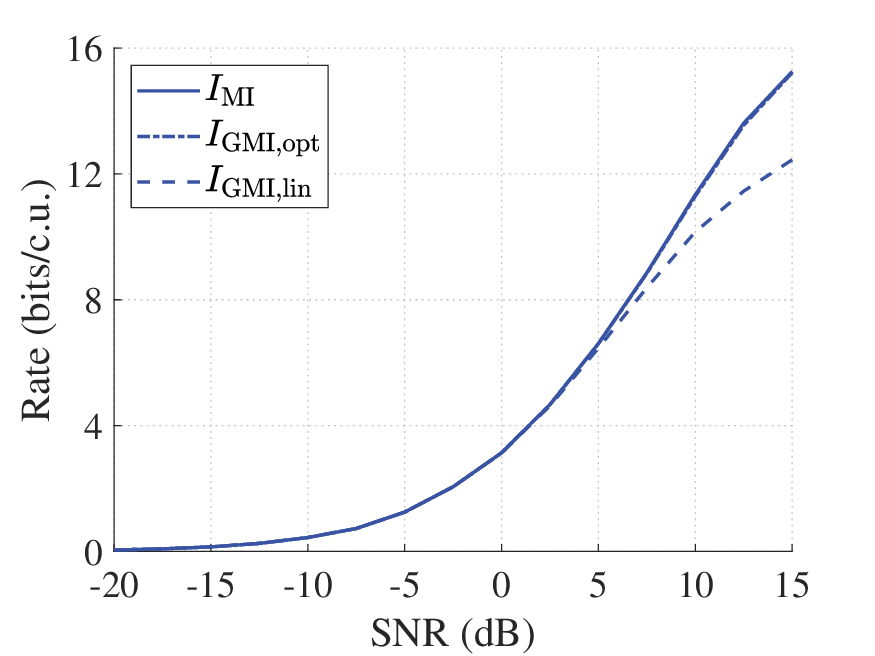}}
\subfigure[$K=8$, $L=8$]{
	\includegraphics[width=0.31\textwidth]
	{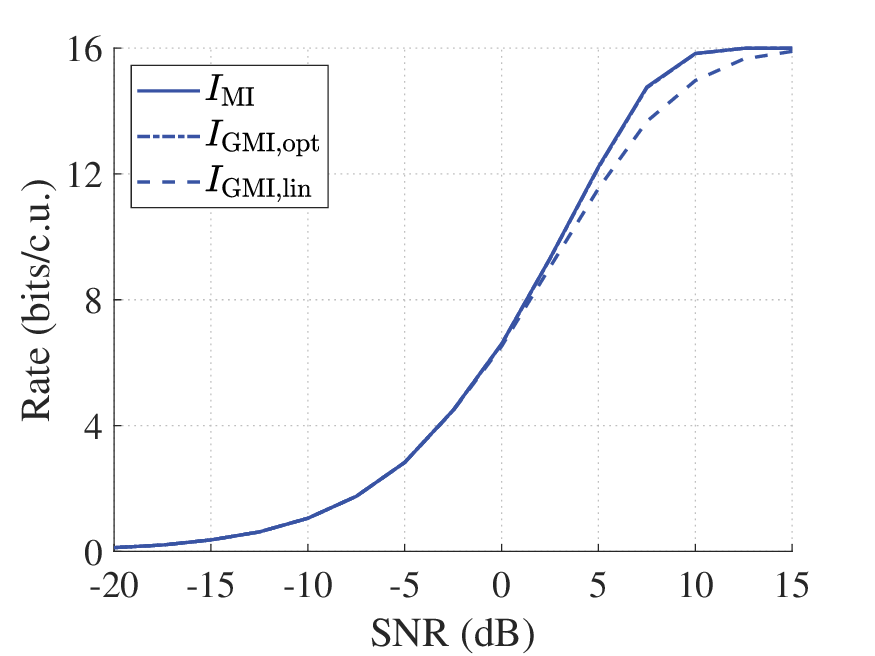}}
\subfigure[$K=8$, $L=16$]{
	\includegraphics[width=0.31\textwidth]
	{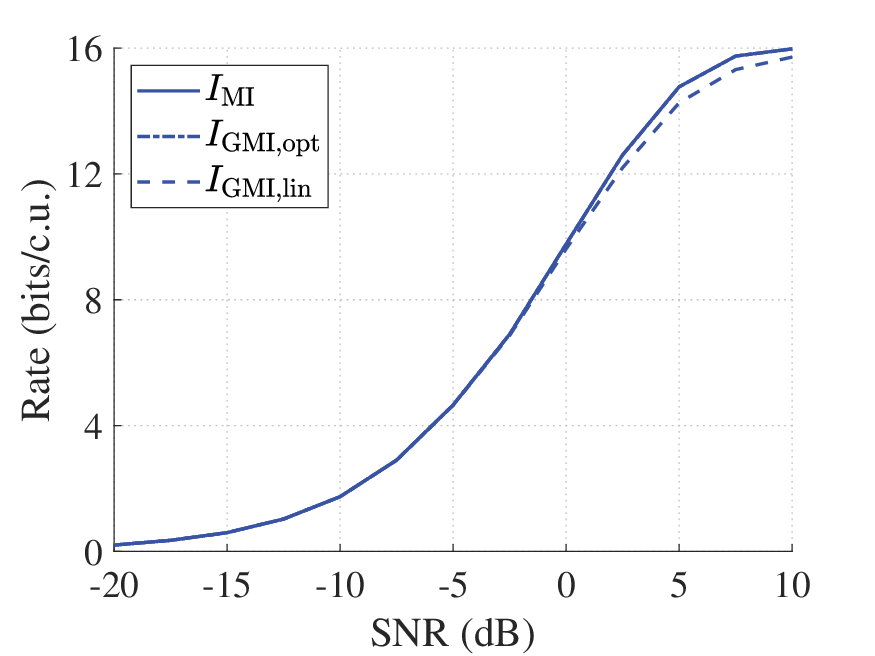}}
	\caption{Sum GMIs and mutual information for multiuser channel \eqref{eqn:gaussian-fading-channel} under GNND with SIC.}
	\label{fig:gmi-perfect-SU-MIOM}
\end{figure*}

For receiver with SIC, we plot in Figure \ref{fig:gmi-perfect-SU-MIOM} the sum GMIs with GNND and CL, respectively, as well as the sum mutual information $I_\mathrm{MI} = I(\rvx_1, \ldots, \rvx_K; \rvy)$, for comparison. Except that the sum rates are improved due to the alleviation of multiuser interference, we observe similar trends as those in Figure \ref{fig:gmi-perfect-MU-MIOM}: the gap between the GMI with GNND and the channel mutual information is nearly invisible; the gap between the GMI with GNND and that with CL is reduced, thanks to the cancellation of decoded interference, but is still non-negligible. So the GNND receiver architecture, again, exhibits an impressive capability of handling non-Gaussian multiuser interference, incurring essentially no rate loss compared with the ML decoder.

\begin{rem}
	Other input constellations besides QPSK can be studied for GNND analogously, and part of the reason for us to focus on QPSK is that it represents a highly discrete input distribution that is perhaps the most ``non-Gaussian'' (i.e., antipodal in each of the in-phase (I) and quadrature (Q) components). As have been seen in Figure \ref{fig:gmi-perfect-MU-MIOM} and Figure \ref{fig:gmi-perfect-SU-MIOM} and will be further seen in Section \ref{sec:coded_modulation}, due to the non-Gaussian characteristic of QPSK, the CL receiver architecture does not effectively handle multiuser interference, while the GNND does so.
\end{rem}

\begin{figure*}[ht]
	\centering
	\subfigure[GNND: $K=8$, $L=16$, $\mathrm{SNR}=9$ dB]{\label{fig:x_hat_perfect_8_16_gnnd}
			\includegraphics[width=0.28\textwidth]
			{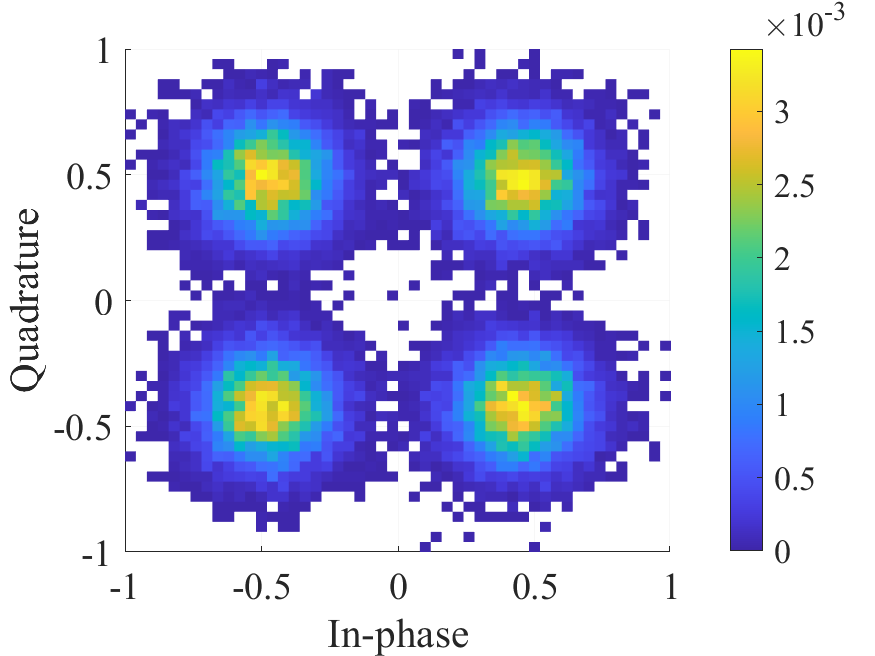}}   
	\subfigure[GNND: $K=8$, $L=8$, $\mathrm{SNR}=15$ dB]{\label{fig:x_hat_perfect_8_8_gnnd}
			\includegraphics[width=0.28\textwidth]
			{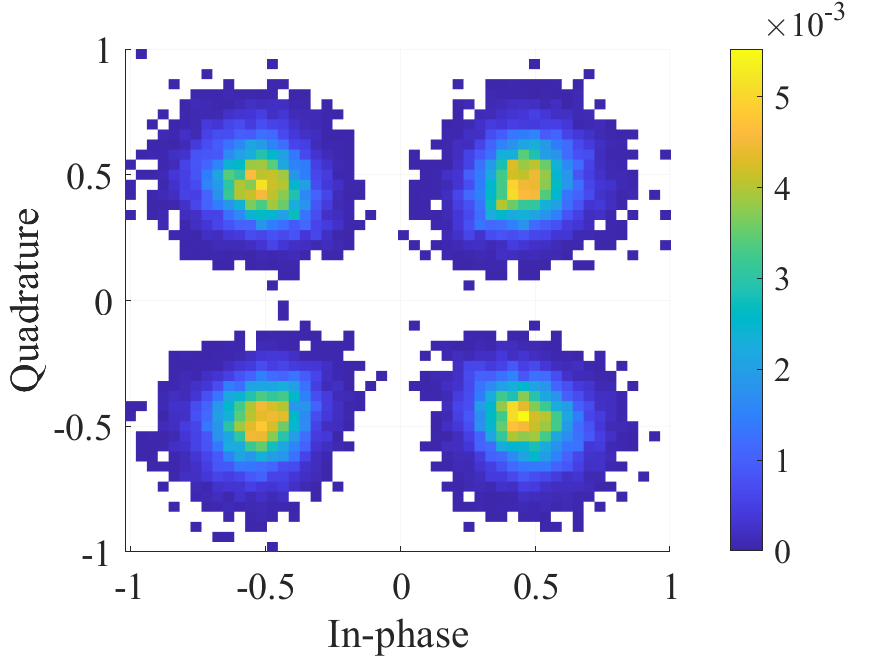}}
	\subfigure[GNND: $K=8$, $L=4$, $\mathrm{SNR}=25$ dB]{\label{fig:x_hat_perfect_8_4_gnnd}
			\includegraphics[width=0.28\textwidth]
			{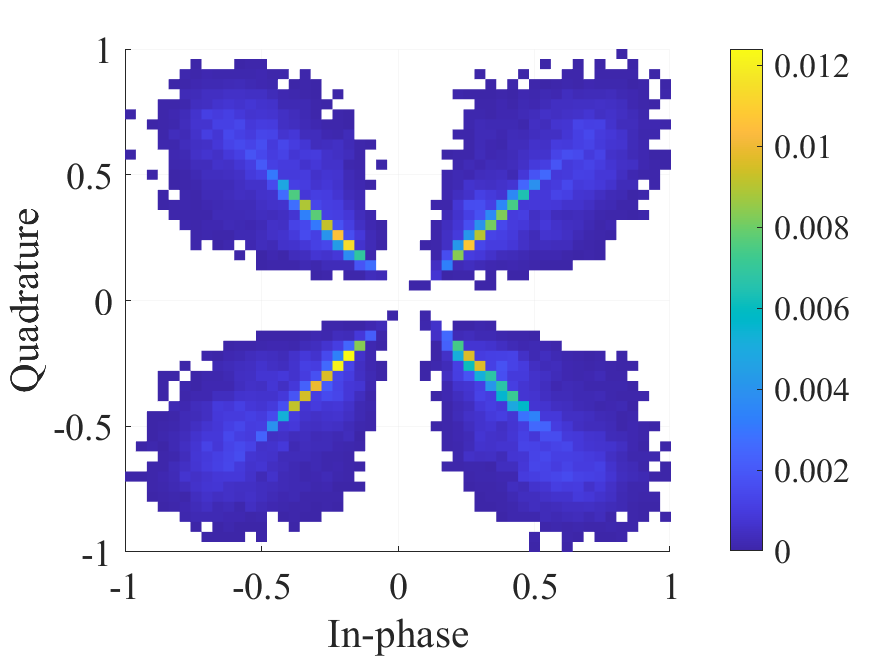}}
	\subfigure[LMMSE: $K=8$, $L=16$, $\mathrm{SNR}=9$ dB]{\label{fig:x_hat_perfect_8_16_lmmse}
			\includegraphics[width=0.28\textwidth]
			{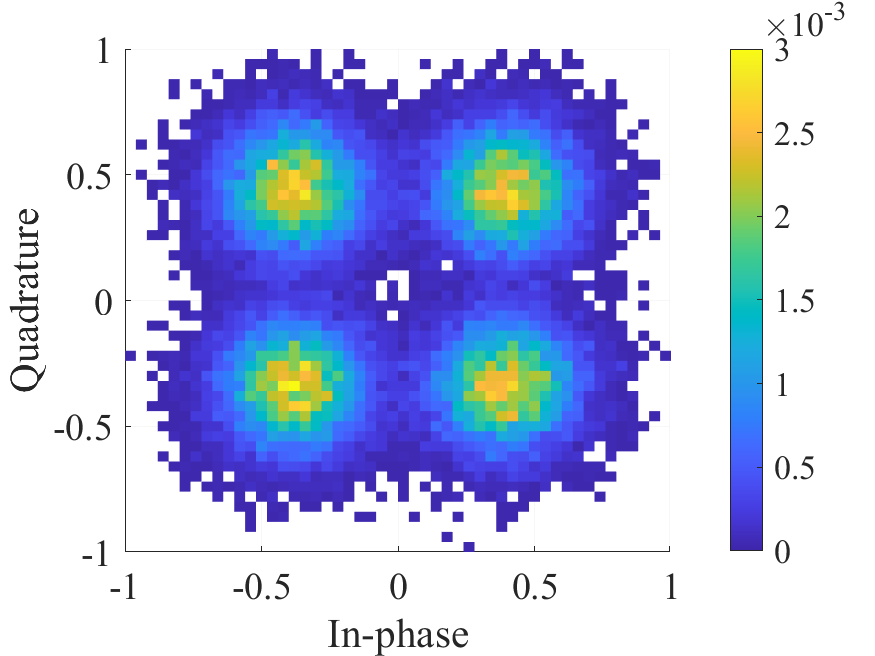}}
	\subfigure[LMMSE: $K=8$, $L=8$, $\mathrm{SNR}=15$ dB]{\label{fig:x_hat_perfect_8_8_lmmse}
			\includegraphics[width=0.28\textwidth]
			{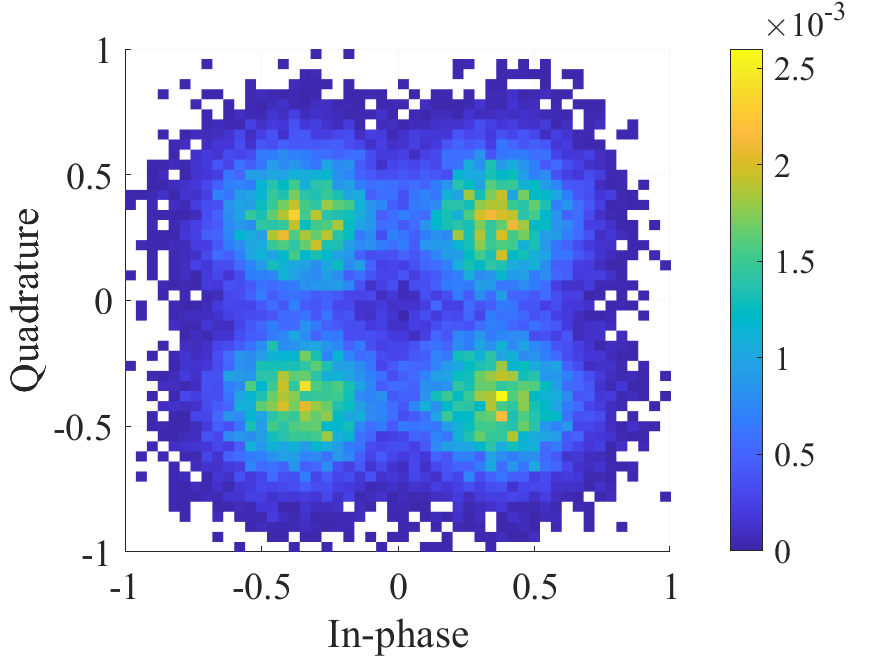}} 
	\subfigure[LMMSE: $K=8$, $L=4$, $\mathrm{SNR}=25$ dB]{\label{fig:x_hat_perfect_8_4_lmmse}
			\includegraphics[width=0.28\textwidth]
			{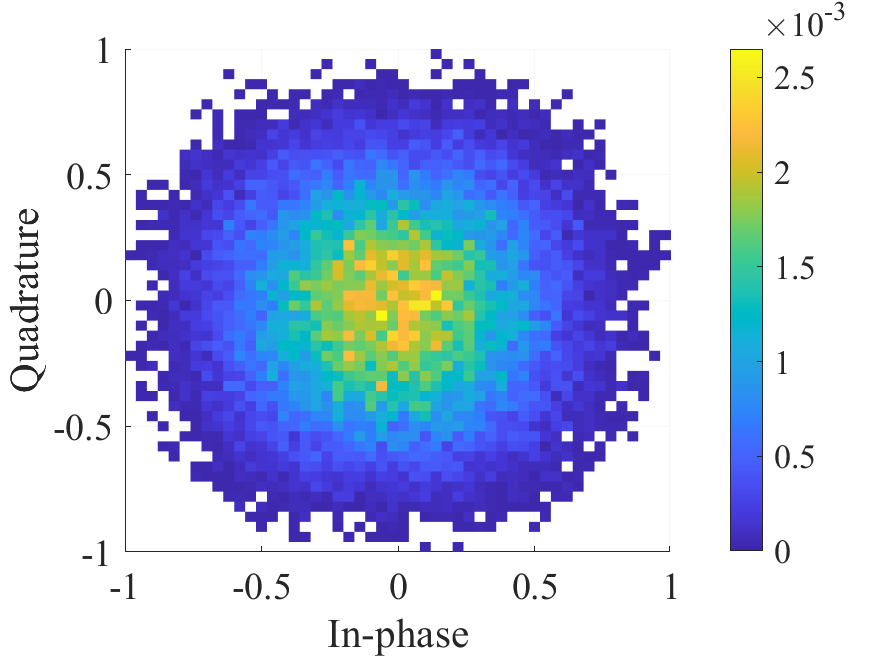}}
	\caption{GNND estimates of $\rvX_1$ and LMMSE estimates of $\rvx_1$, in multiuser channel \eqref{eqn:gaussian-fading-channel}.}
	\label{fig:x-decomposition}
\end{figure*}

To get a heuristic idea about why GNND is so effective, we inspect the GNND estimate of $\rvx_k$ (see the discussion following \eqref{eqn:GNND-qpsk-opt}) and the LMMSE estimate of $\rvx_k$ in CL. The estimates on the I-Q plane are plotted in scattergram, after normalization, in Figure \ref{fig:x-decomposition}. Figures \ref{fig:x_hat_perfect_8_16_gnnd} and \ref{fig:x_hat_perfect_8_16_lmmse} compare the estimates for a under-loaded system ($(K, L) = (8, 16)$); Figures \ref{fig:x_hat_perfect_8_8_gnnd} and \ref{fig:x_hat_perfect_8_8_lmmse} compare the estimates for a fully loaded system ($(K, L) = (8, 8)$); while Figures \ref{fig:x_hat_perfect_8_4_gnnd} and \ref{fig:x_hat_perfect_8_4_lmmse} compare the estimates for an overloaded system ($(K, L) = (8, 4)$). We observe that in all cases, the GNND estimate yields a better resolution among the constellation points, compared with the LMMSE estimate, which typically blurs the decision regions, especially as system load increases.

\section{GNND for Multiuser Interference Suppression: Numerical Experiments}
\label{sec:coded_modulation}

In this section, we further study the performance of GNND for multiuser interference suppression using off-the-shelf codes.

{Inspecting the description of GNND, we see that GNND is not a specific decoding algorithm per se. It is in fact a principle for designing receiver, extending the classical nearest neighbor decoding principle by introducing a symbol-level memoryless preprocessing step (namely processing function $g$ and scaling function $f$). We emphasize that such a preprocessing step can be directly inserted between channel output and off-the-shelf decoder in standard receivers, without any further modification of system. Such a ``plug-and-play'' feature may be desirable when we need to encounter a variety of possible channels in applications, because only the preprocessing step need be adjusted accordingly, while the decoder itself can be kept intact.}

\subsection{An Appetizer: Experiment with Viterbi Decoding of Convolutional Codes}
\label{subsec:viterbi}

In order to verify the benefit of GNND, we begin with a simple setup, where a rate-$1/2$ convolutional code is used. Every information bit produces two coded bits, and such two coded bits are directly mapped to a QPSK constellation point according to the Gray mapping. By modifying the branch metric accordingly, we can use the Viterbi algorithm to genuinely realize GNND, CL, or ML decoding.

\begin{figure*}[ht]
	\centering
	\subfigure[$K=4$, $L=4$, Rate $= 1/2$]{
		\includegraphics[width=0.33\textwidth]
		{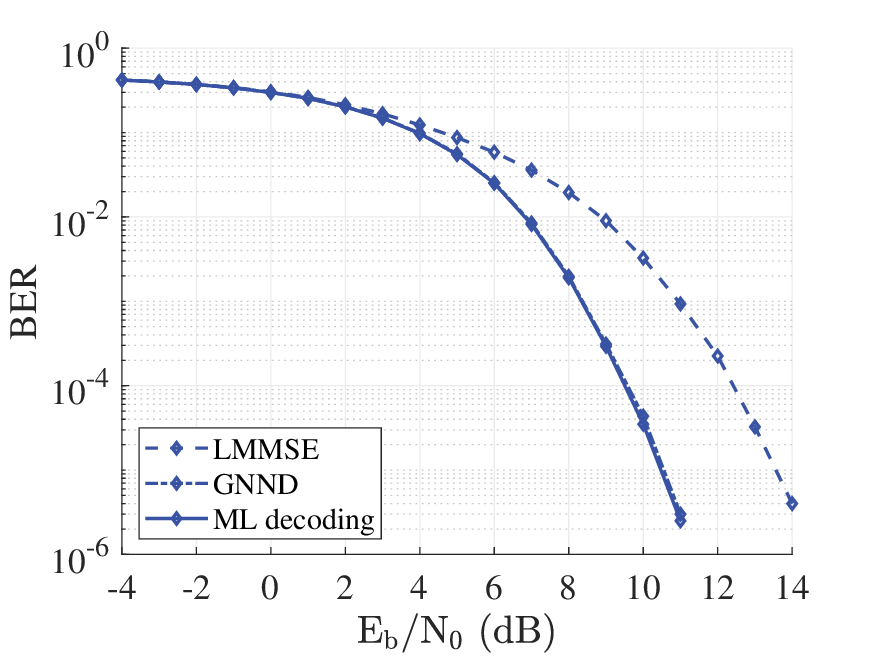}
	}
	\subfigure[$K=8$, $L=8$, Rate $= 1/2$]{
		\includegraphics[width=0.33\textwidth]
		{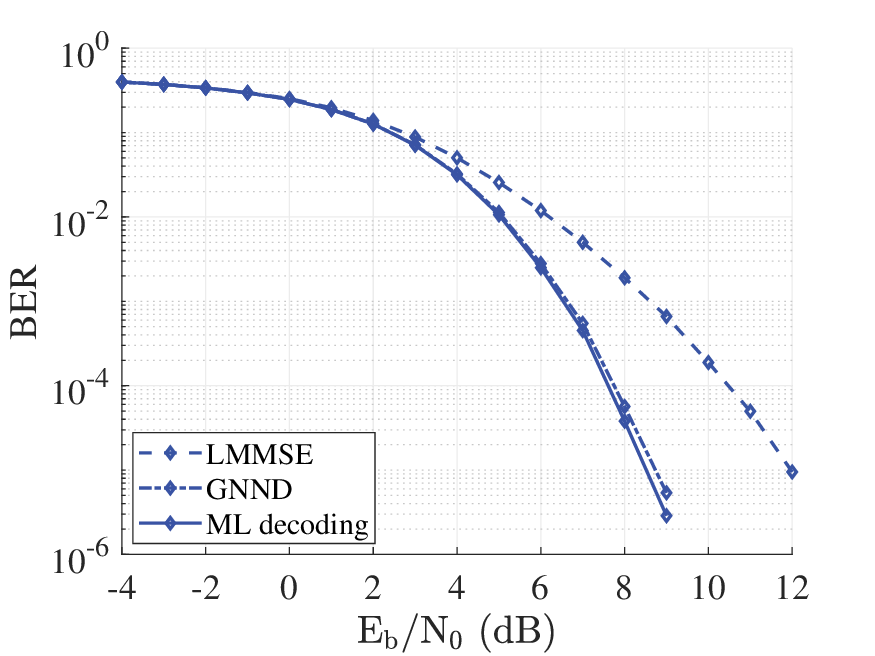}
	}
	\caption{BER performance of the Viterbi algorithm for a rate-$1/2$ convolutional code for a fully loaded system with SIC.}
	\label{fig:viterbi_ber}
\end{figure*}

Figure \ref{fig:viterbi_ber} displays the bit error rate (BER) performance of the Viterbi algorithm, for a fully loaded system ($K = L$) with SIC. The generating polynomials of the convolutional code are chosen as $G^{(1)}(D) = 1 + D^2$ and $G^{(2)}(D) = 1 + D + D^2$. We observe that the gap between GNND and CL is no less than $3$ dB for BER $\leq 10^{-4}$, while the gap between GNND and ML is essentially negligible. Note that such difference in performance stems from the difference among decoding metrics only. So this experiment arguably corroborates the benefit of GNND revealed in our information-theoretic analysis in the previous section.

\subsection{Coded Modulation Scheme}
\label{subsec:cod-mod}

Modern wireless communication systems usually adopt codes more powerful than convolutional codes for high spectral efficiency. For example, LDPC codes have been extensively used in standards \cite{3gpp.38.212}, due to their provably good performance and low-complexity decoding algorithm. We thus proceed to investigate the performance of GNND in a simple LDPC coded modulation scheme, in order to illustrate its potential benefit.

To fix ideas, consider QPSK with the Gray mapping \cite{3gppts38.211}, according to
a binary labeling function $\mu$: $\{0,1\}^2\mapsto \mathcal{X}=\left\{a_{1}, a_{2}, a_{3}, a_{4}\right\}$ which maps blocks of two bits to constellation points. Define the inverse mapping for labeling position $j$ as $\mu^{-1}_j$: $\mathcal{X}\mapsto \{0,1\}$, $j=1,2$; that is, $\mu^{-1}_j(x)$ is the $j$th bit corresponding to constellation point $x$. Accordingly, we define
\begin{equation}
\mathcal{X}_b^j = \{x\in \mathcal{X}: \mu^{-1}_j(x)=b\},\quad j=1,2, \quad b=0,1,
\end{equation}
as the set of constellation points $x$ whose binary labels have value $b \in\{0,1\}$ in their $j$th position. In this work we do not consider more sophisticated coded modulation schemes like bit interleaved coded modulation (BICM), which will be left for future research.

The BP decoding algorithm has been commonly used for LDPC codes. The LLRs of the transmitted symbols are iteratively updated via the BP procedure. For a complex-valued AWGN channel $\rvy = \rvx + \rvz$ where $\rvz$ has variance $\sigma^2$, the LLR is initialized as
\begin{align}
	\mathrm{LLR} &= \log\frac{p (y|\mu^{-1}_j(x) = 0)}{p(y|\mu^{-1}_j(x) = 1)}\nonumber\\
	&= \log\frac{\sum_{x \in \mathcal{X}_0^j} p (y| x)}{\sum_{x \in \mathcal{X}_1^j}p(y| x)} \nonumber\\
	&= \log\frac{\sum_{x \in \mathcal{X}_0^j} e^{-\left|y-x\right|^2/\sigma^2}}{\sum_{x \in \mathcal{X}_1^j} e^{-\left|y-x\right|^2/\sigma^2}}.
	\label{eqn:llr-awgn}
\end{align}

In order to adapt the decoding algorithm for AWGN channels to GNND, we draw an analogy between the nearest neighbor decoding metric $|y - x|^2$ for the AWGN channel and the GNND decoding metric $|g(y) - f(y) x|^2$ in (\ref{eqn:GNND}). Intuitively, this corresponds to the following representation of channel (see also \cite[Fig. 2]{wang22it}):
\begin{equation}
\label{eqn:equivalent channel}
g (\rvy) = f (\rvy) \rvx + \rvu,
\end{equation}
and the GNND aims to find the codeword that minimizes the accumulated norm of $\rvu$ over the coding block. This way, when decoding $\rvx_k$, we initialize the LLR as
\begin{equation}
	\mathrm{LLR}_\mathrm{GNND} = \log\frac{\sum_{x_k \in \mathcal{X}_0^j} e^{-\left|g(y) - f(y) x_k\right|^2/\sigma^2_{\rvu_k}}}{\sum_{x_k \in \mathcal{X}_1^j} e^{-\left|g(y)-f(y) x_k\right|^2/\sigma^2_{\rvu_k}}},
	\label{eqn:llr-nosic}
\end{equation}
where $\sigma^2_{\rvu_k} = \mathbf{E}[|g(\rvy) - f(\rvy) \rvx_k|^2]$. This generic form applies for receiver without SIC, and for receiver with SIC, the previously decoded $(x_1, \ldots, x_{k - 1})$ serves as receive side information, and we need to augment $y$ into $(y, x_1, \ldots, x_{k - 1})$. The processing function $g$ and scaling function $f$ can be obtained according to Theorem \ref{thm:GMI-opt-qpsk}.

On the other hand, the CL adopts the channel representation (\ref{eqn:channel-linearization}); also see Section \ref{subsec:GNND-channel-linearization}:
\begin{equation}
	\rvy = \frac{\mathbf{E}[\rvx^\dag \rvy|v]}{P} \rvx + \rvw(v).
\end{equation}
For receiver without SIC, $v$ is void, and for receiver with SIC, we let $v$ be $(x_1, \ldots, x_{k - 1})$ when considering the decoding of $\rvx_k$. The corresponding LLR can be initialized following the discussion in Section \ref{subsec:GNND-channel-linearization}, i.e., postulating the channel representation as Gaussian and applying noise whitening and maximum ratio combining.

\subsection{Approximate Implementation of GNND}
\label{subsec:approx-mmse}

As shown in Theorem \ref{thm:GMI-general-input} and Theorem \ref{thm:GMI-opt-qpsk}, the implementation of GNND boils down to the computation of the conditional expectation $\mathbf{E}[\rvx_k | y]$ (without SIC) or $\mathbf{E}[\rvx_k | y, x_1, \ldots, x_{k - 1}]$ (with SIC) for user $k$. A direct evaluation of such conditional expectations typically involves numerical multiple integration, and the associated computational cost becomes unacceptable as system scales.

To address this challenge, the key is to recall the well known fact that the conditional expectation operator is exactly the MMSE estimator, minimizing the quadratic loss of estimation. Such conditional expectation operator has played a central role in statistical regression problems, and is amenable to machine learning techniques (see, e.g., \cite{hastie}). Therefore, we adopt a neural network for approximating the conditional expectation operator, and this leads to an approximate implementation of GNND.

We stress that, the neural network approximation of the conditional expectation operator only corresponds to a module in GNND, while the remaining computation of GNND and the subsequent BP-based iterative decoding are left unaffected. This is in sharp contrast to end-to-end learning based transceivers (e.g., \cite{oshea,park20icassp,aoudia22twc,raviv23twc}). Furthermore, training neural networks subject to a quadratic loss has been a fairly standard machine learning task, and can be efficiently and stably realized with a standard neural network architecture and a tiny size of parameters, as will be described subsequently. {In contrast, if one insists on implementing a neural network approximation of ML decoding,} the task would then boil down to obtaining a neural network based estimate of the channel input-output conditional probability distribution, i.e., conditional density estimation, which has been a notoriously challenging problem typically requiring an excessive amount of training data and sophisticated neural network design \cite{rothfuss}.

We also note that, in light of the information-theoretic analysis conducted thus far, the neural network approximation considered here is highly interpretable. That is, as long as the neural network approximation satisfactorily yields an accurate estimate of the conditional expectation operator, the resulting approximate implementation of GNND, according to Section \ref{sec:rate_analysis}, would achieve a GMI close to the mutual information. This interpretability is attractive and in sharp contrast to black-box machine learning techniques which are heuristic in nature. It should be noted that, the estimated channel gain coefficients are used only for generating the training dataset, while the model training process does not require any knowledge of the channel model; --- in fact, even the expression of (\ref{eqn:gaussian-fading-channel}) is unnecessary there.

The neural network is deployed as part of the decoder. In order to train it, the decoder needs to generate a training dataset, i.e., a set of i.i.d. channel input-output pairs like $([\rvx_1, \ldots, \rvx_K], \rvy)$, according to the multiuser channel model (\ref{eqn:gaussian-fading-channel}). This can be readily accomplished given the knowledge of the channel gain coefficients $[\bm h_1, \ldots, \bm h_K]$. In practice, when such knowledge is not available, one can employ a channel estimator to obtain estimated (and hence generally imperfect) channel gain coefficients $[\hat{\bm h}_1, \ldots, \hat{\bm h}_K]$. Since the focus of the present work is on GNND, we will not elaborate upon the design of channel estimator here. It suffices to remark that depending upon channel modeling assumptions there are a variety of channel estimators for choice, including the least square (LS) estimator which requires the essentially no prior statistical knowledge of channel, the linear MMSE estimator which requires a prior probability distribution of $[\bm h_1, \ldots, \bm h_K]$, estimators based upon certain sparsity assumptions of $[\bm h_1, \ldots, \bm h_K]$, neural network based estimators, and so on. {In the numerical experiments, we adopt the linear MMSE estimator.}

Based upon the preceding considerations, in this work we use a standard fully-connected neural network. As will be illustrated in the numerical experiments, such a simple design largely suffices for our purpose; more sophisticated designs are potential research topic for future study. For completeness we list some key aspects about our neural network design as follows.

\textit{Structure:} The designed fully-connected neural network has several dense stacks of layers, including hidden layers and one linear output layer. The activation function of each hidden layer is a Rectified Linear Unit (ReLU), defined as
\begin{equation}
f_\mathrm{ReLU}(x) = \max (0,x).
\end{equation}

{\textit{Training set generation and preprocessing:}} In order to train the neural network model, a set of $T$ i.i.d. training samples in the form of $([\rvx_1, \ldots, \rvx_K], \rvy)$ need be generated. An input sample $[\rvx_1, \ldots, \rvx_K]$ is generated according to the specified channel input constellation, and the corresponding output sample $\rvy$ is generated according to the multiuser channel (\ref{eqn:gaussian-fading-channel}), {wherein the channel gain coefficients $[\bm h_1, \ldots, \bm h_K]$ are replaced by their estimates $[\hat{\bm h}_1, \ldots, \hat{\bm h}_K]$.} In order to render the complex-valued samples compatible with existing neural network implementation, a preprocessing step of vectorization is applied to represent the training set as a concatenation of real and imaginary components.

{\textit{Model training:}} As our purpose is to estimate $\mathbf{E}[\rvx_k | y]$ (without SIC) or $\mathbf{E}[\rvx_k | y, x_1, \ldots, x_{k - 1}]$ (with SIC) for each user $k = 1, \ldots, K$, we train the neural network using the quadratic loss between the neural network output $\hat{x}_k$ and the corresponding channel input $x_k$; that is, for each user $k$, we construct a neural network model and minimize the loss function
\begin{equation}
f_{k, \mathrm{Loss}} = \frac{1}{T} \sum_{t=1}^T\left|\hat x_{k,t} - x_{k,t}\right|^2, 
\end{equation}
where $\hat x_{k, t}$ is the neural network output corresponding to the $t$-th input sample. We employ the mini-batch gradient descent algorithm and the training is implemented using the Adaptive Moment Estimation (Adam) optimizer \cite{kingma15iclr}. The thus trained model approximately implements the conditional expectation operator, in other words, the MMSE estimator, denoted by $\hat{\mathbf{E}}[\rvx_k | y]$ (without SIC) or $\hat{\mathbf{E}}[\rvx_k | y, x_1, \ldots, x_{k - 1}]$ (with SIC), for each user $k = 1, \ldots, K$, which can then be fed into the expression of GNND for initializing the LLRs according to the description in Section \ref{subsec:cod-mod}. It is clear that the model training process does not need the knowledge of the channel model (\ref{eqn:gaussian-fading-channel}) at all.

We briefly remark on the computational cost. In the approximate implementation of GNND, we use a fully connected neural network with a fixed number of layers, and its width is equal to $2L$ to account for real and imaginary parts of the received signal. With $K$ users, we need to invoke the neural network $K$ times. Therefore, the computational complexity is proportional to $KL$. In contrast, for LMMSE in CL, each user needs to compute the inverse of a matrix of dimension $L$, and hence the overall computational complexity is proportional to $KL^3$ for successively decoding $K$ users; for ML, the computational complexity is further exponentially increasing with $K$.

\begin{figure*}[ht]
	\centering
	\subfigure[GNND: perfect CSI]{\label{fig:x_hat_imperfect_8_16_gnnd_infP}
		\includegraphics[width=0.23\textwidth]
		{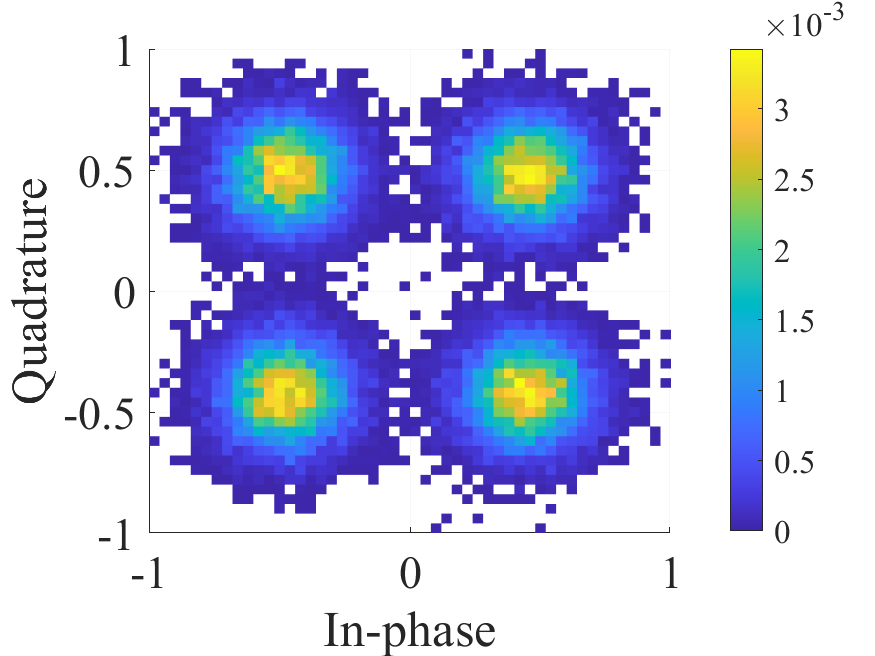}}
	\subfigure[GNND: $|x_{\mathrm{p}}|^2=16P$]{\label{fig:x_hat_imperfect_8_16_gnnd_16P}
		\includegraphics[width=0.23\textwidth]
		{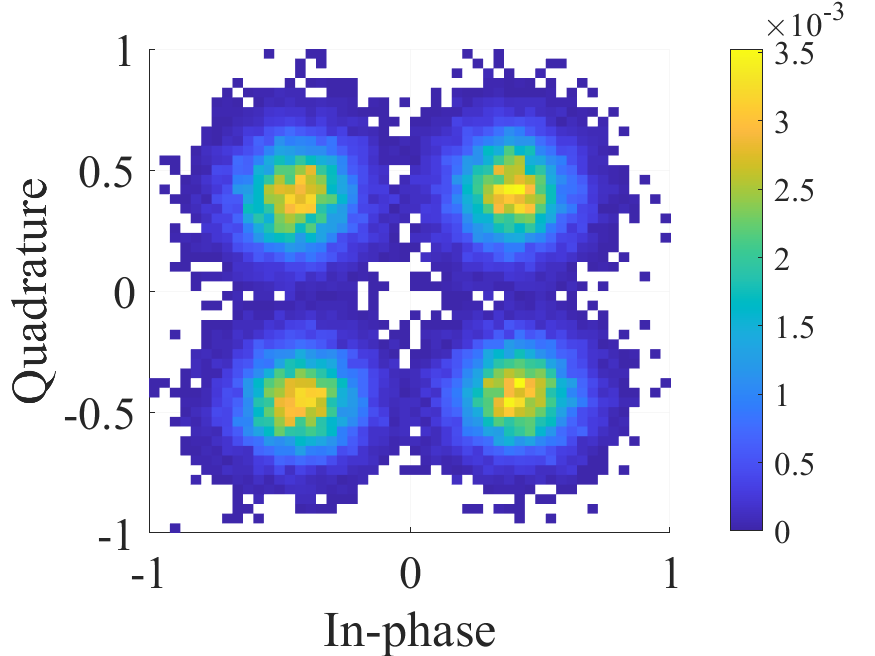}}
	\subfigure[GNND: $|x_{\mathrm{p}}|^2=4P$]{\label{fig:x_hat_imperfect_8_16_gnnd_4P}
		\includegraphics[width=0.23\textwidth]
		{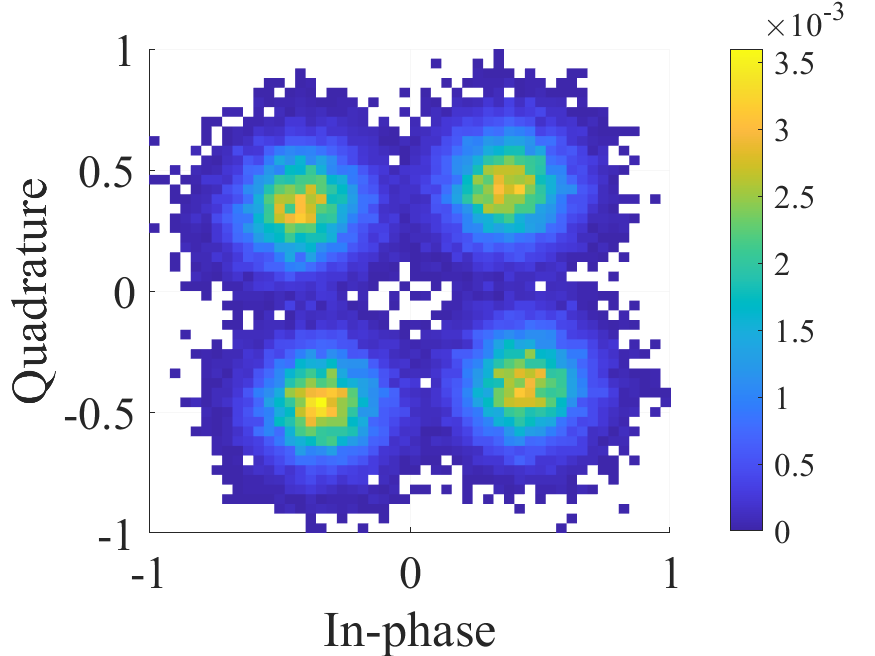}}
	\subfigure[GNND: $|x_{\mathrm{p}}|^2=P$]{\label{fig:x_hat_imperfect_8_16_gnnd_1P}
		\includegraphics[width=0.23\textwidth]
		{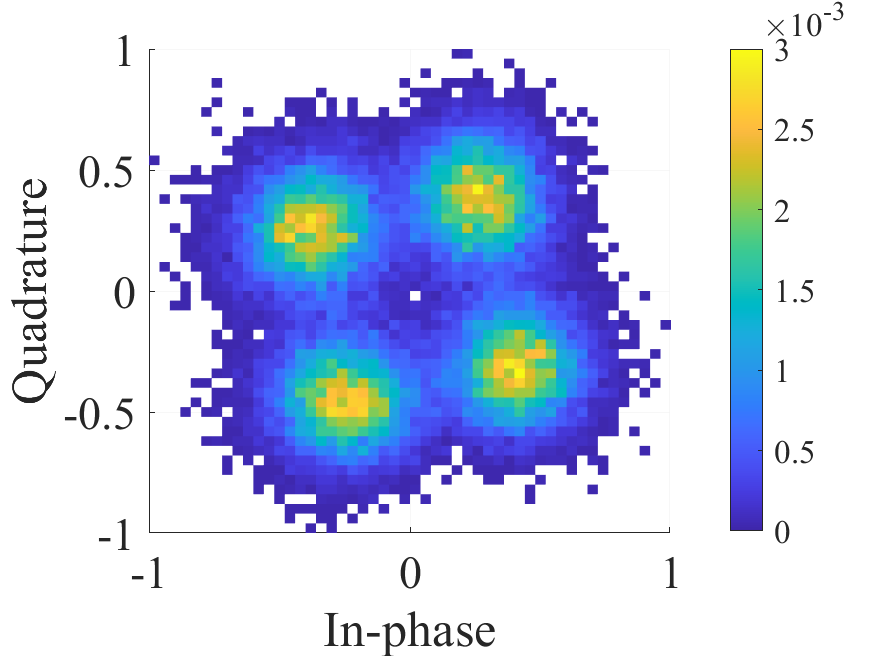}}
	\subfigure[LMMSE: perfect CSI]{\label{fig:x_hat_imperfect_8_16_lmmse_infP}
		\includegraphics[width=0.23\textwidth]
		{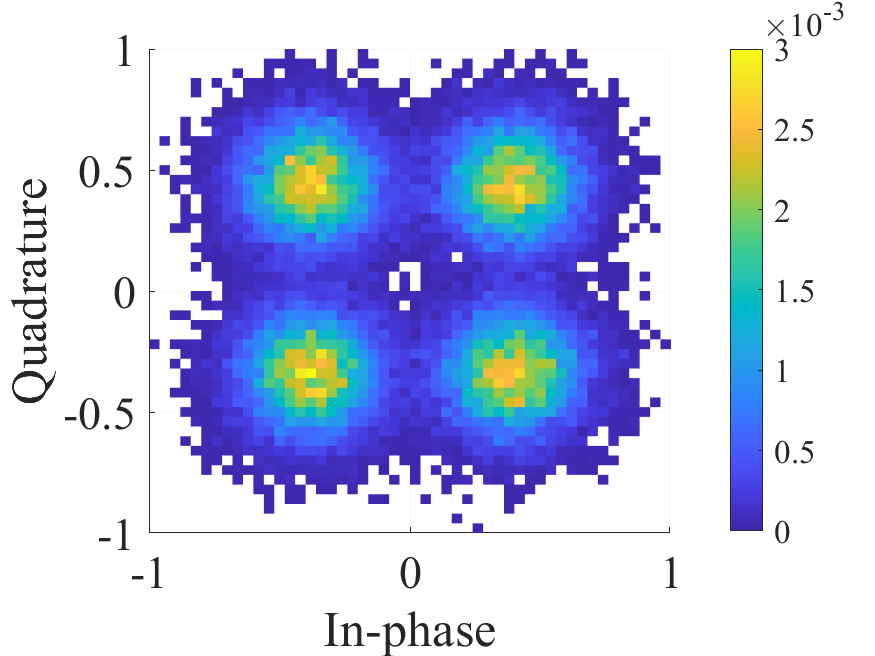}}
	\subfigure[LMMSE: $|x_{\mathrm{p}}|^2=16P$]{\label{fig:x_hat_imperfect_8_16_lmmse_16P}
		\includegraphics[width=0.23\textwidth]
		{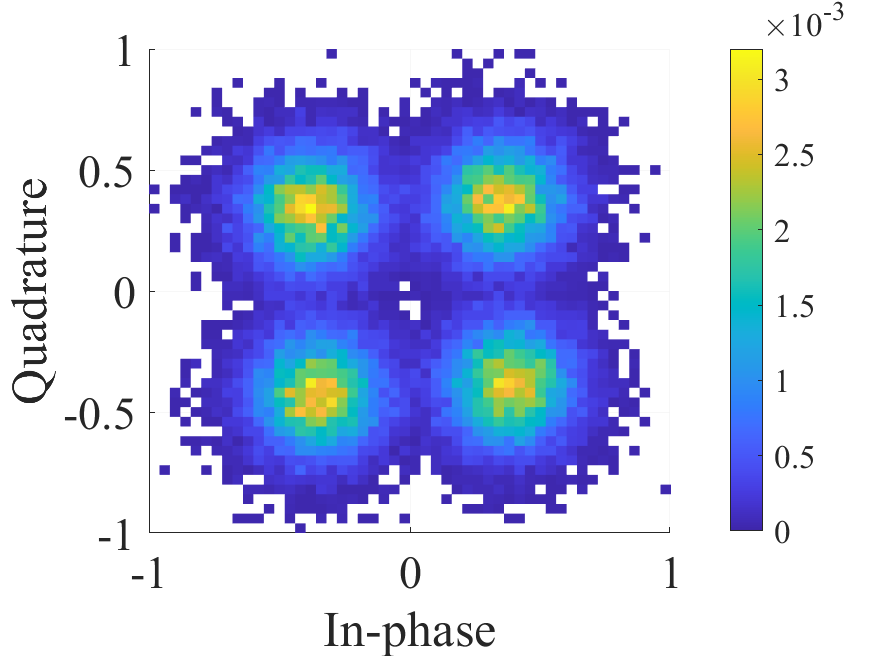}}
	\subfigure[LMMSE: $|x_{\mathrm{p}}|^2=4P$]{\label{fig:x_hat_imperfect_8_16_lmmse_4P}
		\includegraphics[width=0.23\textwidth]
		{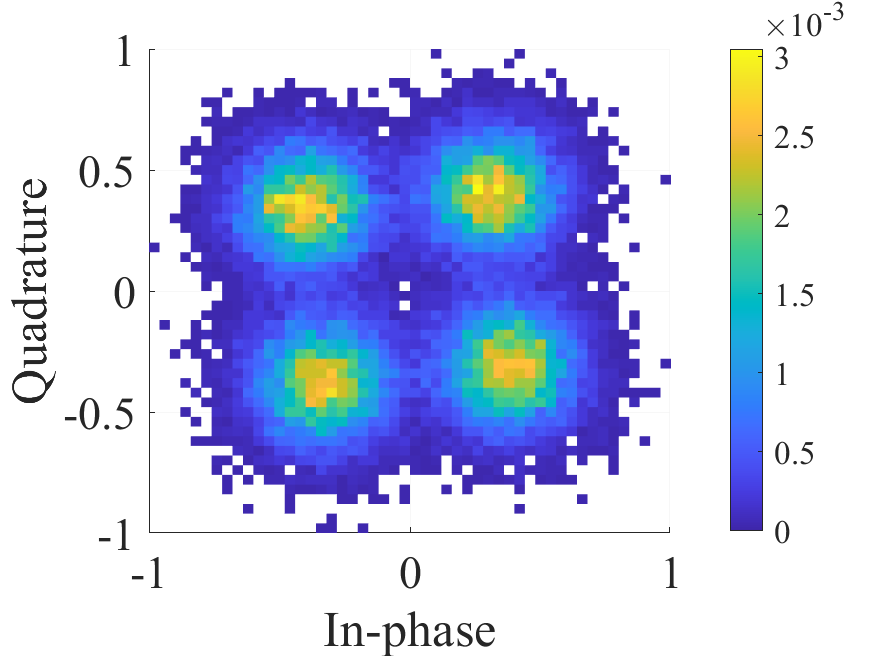}}
	\subfigure[LMMSE: $|x_{\mathrm{p}}|^2=P$]{\label{fig:x_hat_imperfect_8_16_lmmse_1P}
		\includegraphics[width=0.23\textwidth]
		{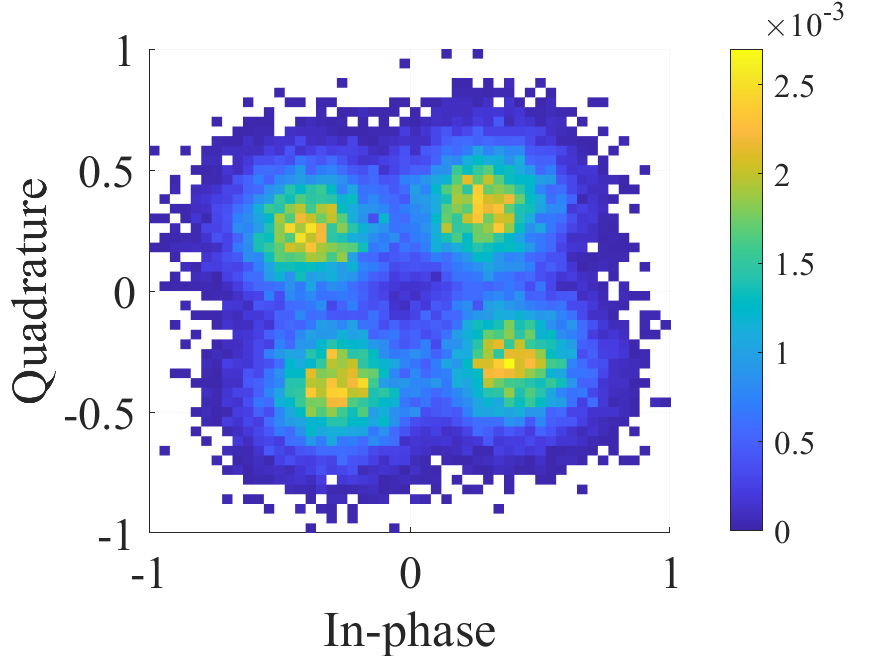}}
	\caption{GNND estimates of $\rvX_1$ when GNND is approximately implemented using neural networks, and LMMSE estimates of $\rvx_1$, in multiuser channel \eqref{eqn:gaussian-fading-channel}, under different pilot power levels $|x_{\mathrm{p}}|^2$ (perfect channel state information (CSI) corresponds to the limiting case where the pilot power level  $|x_{\mathrm{p}}|^2$ is infinite), for $K=8$, $L=16$, $\mathrm{SNR}=9$ dB.} 
	\label{fig:x-decomposition-learned}
\end{figure*}

\subsection{Numerical Results}
\label{subsec:numerical-results}

The numerical results presented in this subsection is for the multiuser channel model \eqref{eqn:gaussian-fading-channel} with the number of users $K \in \{8, 16\}$ and the number of receive antennas $L \in \{8, 16, 32\}$. We consider receiver without SIC, a common choice in many practical wireless systems with such a moderate number of users. Quasi-static i.i.d. Rayleigh fading is assumed, and the channel gain coefficients are obtained via pilots.

As outlined in Section \ref{subsec:cod-mod}, we use QPSK modulation with the Gray mapping to implement the coded modulation scheme. The channel code is a standard quasi-cyclic LDPC code whose parity check matrix is constructed following the specification given in \cite{3gpp.38.212}, decoded by the standard BP algorithm.

For model training, we follow the approach described in Section \ref{subsec:approx-mmse}, generating the training dataset based upon the estimated channel gain coefficients in the channel model \eqref{eqn:gaussian-fading-channel}. For each user we build a fully-connected neural network whose three hidden layers have 200, 100, and 50 neurons respectively, and whose output layer consists of two neurons to output the real and imaginary components of the conditional expectation respectively. The input dimension of the neural network depends upon the number of users and the number of receive antennas. From the perspective of modern deep learning, such a neural network is tiny, but it suffices to yield significant performance gain compared with the CL approach, as will be presented in the sequel.

We train the neural network with 400000 samples for 100 epochs, with 200 batches and  2000 samples in each batch. The learning rate is chosen as $\lambda = 0.001$.

Before presenting the BER performance, similar to Figure \ref{fig:x-decomposition}, we plot in Figure \ref{fig:x-decomposition-learned} the scattergram of the estimates of $\rvx_1$ (a typical user) on the I-Q plane, when the GNND estimators are approximately implemented using neural networks. Comparing the scattergram plots under different pilot power levels, we observe the impact of imperfect channel knowledge from its resulting phase rotation effect; comparing the scattergram plots for GNND and LMMSE estimates, we find that GNND estimates, even approximately implemented, still yields visibly better resolution among the constellation points.

\begin{figure*}[ht]
	\centering
	\subfigure[$K=8$, $L=8$, Rate $= 5/6$]{
\includegraphics[width=0.35\textwidth]
{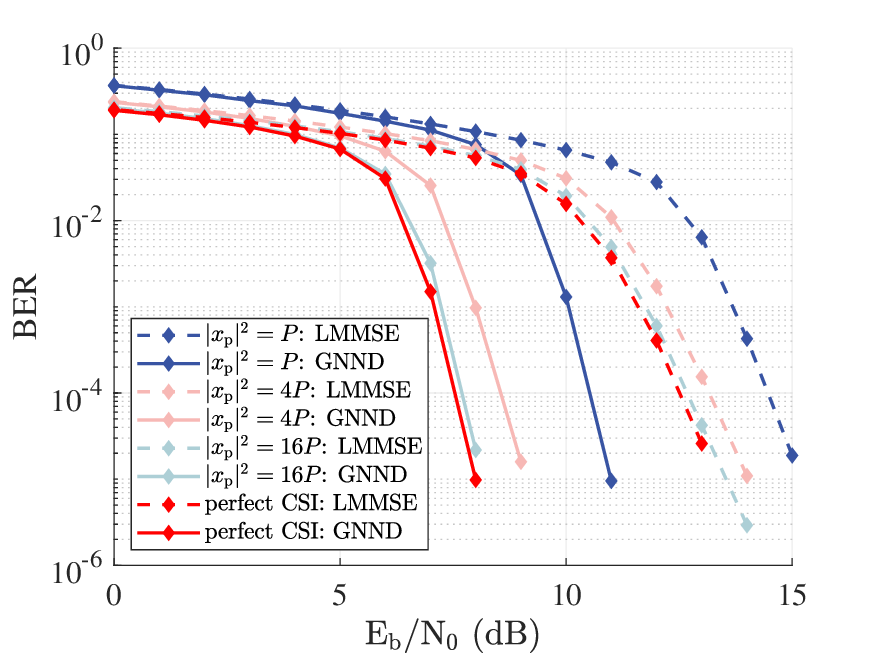}
	}
    \subfigure[$K=8$, $L=16$, Rate $= 5/6$]{
\includegraphics[width=0.35\textwidth]
{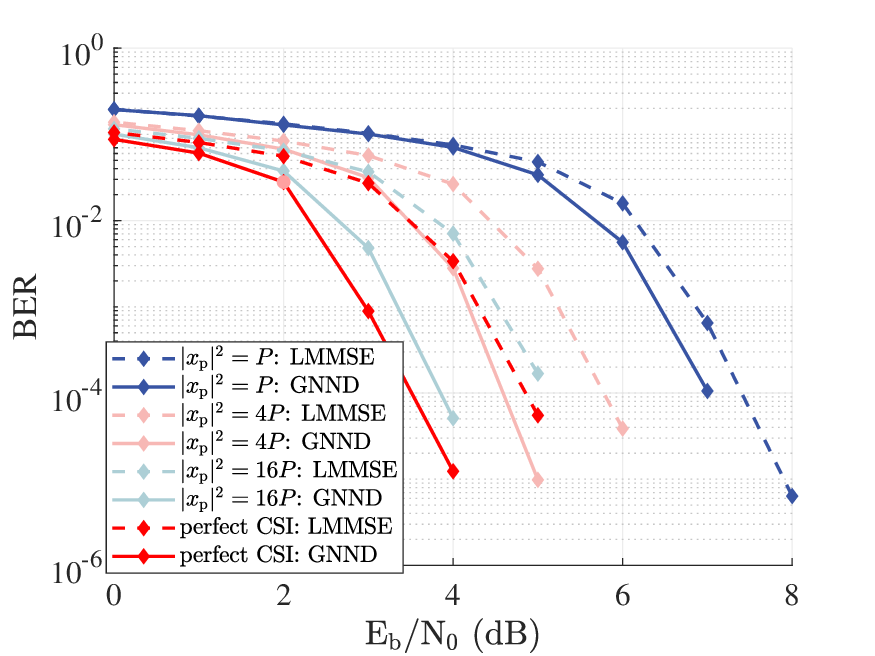}
	}
     \subfigure[$K=16$, $L=16$, Rate $= 5/6$]{
\includegraphics[width=0.35\textwidth]
{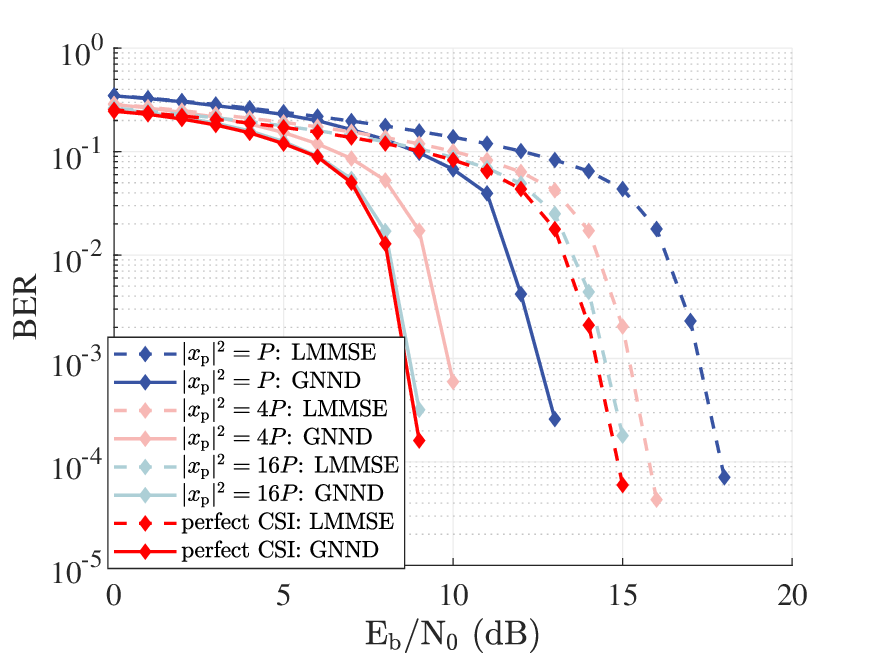}
	}
      \subfigure[$K=16$, $L=32$, Rate $= 5/6$]{
\includegraphics[width=0.35\textwidth]
{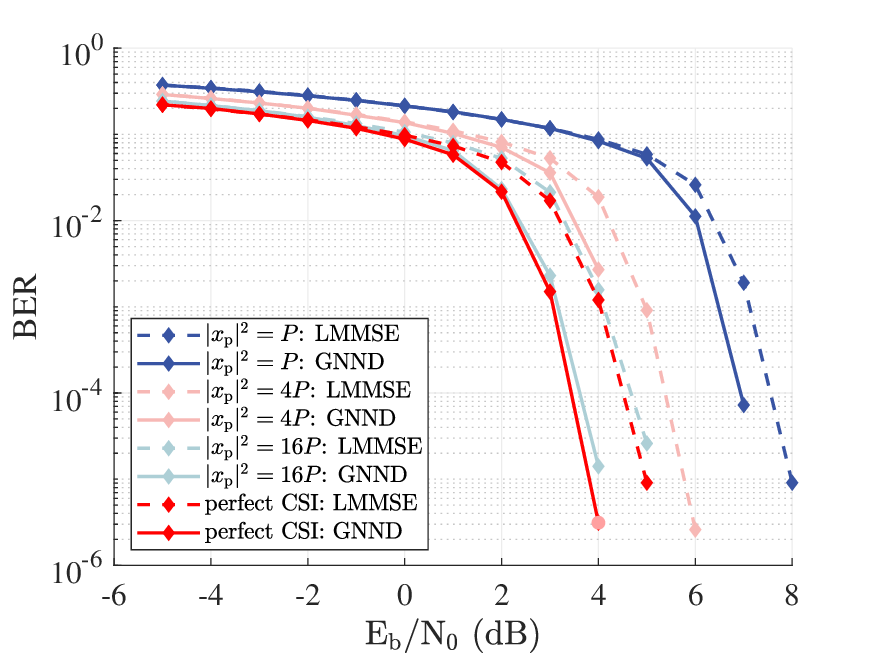}
	}
\caption{BER performance of the LDPC coded modulation scheme.}
\label{fig:ber-w-training-w-nn}
\end{figure*}

We plot in Figure \ref{fig:ber-w-training-w-nn} the BER performance of GNND and CL. The information bit length is $440$ and the code rate is $5/6$ for each user. As shown via the information-theoretic analysis in Section \ref{sec:rate_analysis}, the benefit of GNND becomes evident as the system load (i.e., $K/L$) increases. Therefore, here we present the numerical results for half-loaded ($(K, L) = (8, 16)$ or $(16, 32)$) or fully-loaded ($(K, L) = (8, 8)$ or $(16, 16)$) configurations. We do not present results for overloaded configurations because then the CL approach performs poorly. For each configuration, we plot the BER curves of GNND (approximately implemented via neural networks) and CL, under different pilot power levels. We observe that there exists a multi-dB gap between GNND and CL, for each configuration and each pilot power level. This demonstrates the consistent performance gain of GNND over CL, corroborating the information-theoretic analysis in Section \ref{sec:rate_analysis}. Different pilot power levels correspond to different levels of channel estimation quality, and we observe that, as the pilot power level increases, the BER performance quickly approaches that of the ideal case of perfect knowledge of channel. This suggests that it is sensible to employ a peaky pilot for channel training, as is already widely adopted in practice.

\section{Conclusion}

\label{sec:conclusion}

GNND is a conceptually simple approach for increasing the achievable information rate and improving the decoding performance in channels where nearest neighbor decoding is no longer optimal. Via appropriate output processing and codeword scaling, the resulting GNND achieves higher GMI compared with the conventional, yet commonly adopted, CL approach, while still retains an additive noise channel representation, which renders GNND seamlessly compatible with Gaussian channel-based decoders.

Under general input constellation, we have derived the criteria of the optimal GNND, which possesses an interesting and insightful conditional moments matching interpretation. Under simple constellations such as QPSK, the decoding metric of the optimal GNND and its corresponding GMI can be further derived in closed form. An important task for future research is developing efficient algorithms for solving the optimal GNND under more sophisticated constellations.

On the implementation side, a main challenge for GNND is the evaluation of conditional expectation operator. We have attempted to employ a neural network based approximate implementation and demonstrated its effectiveness for the case study of multiuser interference suppression. For simplicity, we have separated channel training and GNND model training. {In principle, if computational capability is abundant, the channel training step and the subsequent GNND model training step may be integrated together within a single large-scale neural network.} {For implementing the neural network based approximation of the conditional expectation operator in practical wireless communication systems, the time and storage overhead for model training needs to be considered. Further investigation on online model training, especially in time-varying channels where the requirement on time overhead is stringent, and in massive-antenna channels where the requirement on storage overhead is demanding, is an interesting future research topic.}

{\appendices

\section{Proof of Theorem \ref{thm:GMI-general-input}}
\label{sec:proof of theorem 1}
For simplicity of exposition, and in light of Remark \ref{rem:augmented-output}, throughout the following proofs we use $\rvy$ to represent the augmented channel output $(\rvy, \rvv)$, when $\rvv$ is present.

With the decoding metric $d(x, y) = |g(y) - f(y)x|^2$, we inspect the resulting general expression of the GMI \eqref{eqn:gmi-general-constellation}. For optimizing $I_{\mathrm{GMI},g,f}$ over $(g, f)$, noting that the optimality will not be lost with scaling by a common constant, as remarked following (\ref{eqn:decoder-GNND-optimal-g}) and (\ref{eqn:decoder-GNND-optimal-f}), we may eliminate the variable $\theta < 0$ by redefining $g$ as $\sqrt{-\theta} g$ and $f$ as $\sqrt{-\theta} f$ respectively. So we have the optimization problem (\ref{eqn:optimization-GMI-gf}). Note that here and in the sequel $\rvy$, $y$ and $\mathcal{Y}$ should be understood as $(\rvy, \rvv)$, $(y, v)$ and $\mathcal{Y} \times \mathcal{V}$, respectively, whenever applicable.
\begin{figure*}
\begin{equation}\label{eqn:optimization-GMI-gf}
    \sup_{g, f} I_{\mathrm{GMI},g,f} = \sup_{g, f} \left\{- \sum_{(x, y) \in \mathcal{X}\times \mathcal{Y}} p_\rvx(x) p_{\rvy|\rvx}(y|x) |g(y) - f(y) x|^2 \right.
    \left. - \sum_{y \in \mathcal{Y}} p_\rvy(y) \log \sum_{x' \in \mathcal{X}} p_\rvx(x') e^{- |g(y) - f(y) x'|^2} \right\}.
\end{equation}
\end{figure*}

Applying the law of total expectation, and noting that both $g$ and $f$ are functions of $y$, we can rewrite (\ref{eqn:optimization-GMI-gf}) as (\ref{eqn:optimization-GMI-gf-decomposed}).
\begin{figure*}
\begin{equation}\label{eqn:optimization-GMI-gf-decomposed}
    \sup_{g, f} I_{\mathrm{GMI},g,f} = - \sum_{y \in \mathcal{Y}} p_\rvy(y) \inf_{g(y), f(y)} \left\{ \sum_{x \in \mathcal{X}} p_{\rvx|\rvy}(x|y) |g(y) - f(y) x|^2 + \log \sum_{x' \in \mathcal{X}} p_\rvx(x') e^{- |g(y) - f(y) x'|^2} \right\}.
\end{equation}
\end{figure*}
This way, the original optimization over functions $(g, f)$ is decomposed into a collection of finite-dimensional optimization sub-problems over variables $(g(y), f(y))$, indexed by $y \in \mathcal{Y}$. A similar technique has been adopted in \cite[Thm. 1]{wang22it}.

Now for each $y \in \mathcal{Y}$, we expand the objective function inside the bracket in (\ref{eqn:optimization-GMI-gf-decomposed}) as (\ref{eqn:obj-y}).
\begin{figure*}
\begin{align}
    \label{eqn:obj-y}
    & \quad\sum_{x \in \mathcal{X}} p_{\rvx|\rvy}(x|y) |g(y) - f(y) x|^2 + \log \sum_{x' \in \mathcal{X}} p_\rvx(x') e^{- |g(y) - f(y) x'|^2}\nonumber\\
    &= |g(y)|^2 + |f(y)|^2 \sum_{x \in \mathcal{X}} |x|^2 p_{\rvx|\rvy}(x|y) - g(y)f(y)^\dag \sum_{x\in \mathcal{X}} x^\dag p_{\rvx|\rvy}(x|y) - g(y)^\dag f(y) \sum_{x\in \mathcal{X}} x p_{\rvx|\rvy}(x|y) \nonumber\\
    &\quad\quad\quad\quad - |g(y)|^2 + \log \sum_{x' \in \mathcal{X}} p_\rvx(x') e^{-|f(y)|^2 |x'|^2 + g(y)f(y)^\dag x'^\dag + g(y)^\dag f(y) x'}\nonumber\\
    &= |f(y)|^2 \mathbf{E}[|\rvx|^2|y] - g(y) f(y)^\dag \mathbf{E}[\rvx|y]^\dag - g(y)^\dag f(y) \mathbf{E}[\rvx|y] + \log \mathbf{E}\left[e^{-|f(y)|^2 |\rvx|^2 + g(y) f(y)^\dag \rvx^\dag + g(y)^\dag f(y) \rvx}\right].
\end{align}
\end{figure*}

With some algebraic manipulations and via introducing $\alpha = \Re\{g^\dag(y)f(y)\}$, $\beta = \Im\{g^\dag(y)f(y)\}$, and $\gamma = |f(y)|^2$, (\ref{eqn:obj-y}) becomes
\begin{align}
    \label{eqn:obj-y-new}
    &\gamma \mathbf{E}[|\rvx|^2|y] - 2\alpha\Re\{ \mathbf{E}[\rvX|y]\}+2\beta\Im\{ \mathbf{E}[\rvX|y]\} \nonumber\\
    & \quad\quad\quad\quad + \log \mathbf{E} \left[ e^{-\gamma|\rvx|^2 + 2\alpha \Re\{\rvx\} - 2\beta \Im\{\rvx\}}\right].
\end{align}

Letting the partial derivatives of (\ref{eqn:obj-y-new}) with respect to $\alpha$, $\beta$, and $\gamma$ all vanish, we obtain
\begin{align}
    \label{eqn:alpha-beta-gamma-1st}
    \mathbf{E}[\rvX|y] &= \frac{\mathbf{E} \left[\rvx e^{-\gamma|\rvx|^2+2\alpha\Re\{\rvx\}-2\beta\Im\{\rvx\}}\right]}{\mathbf{E} \left[ e^{-\gamma|\rvx|^2+2\alpha\Re\{\rvx\}-2\beta\Im\{\rvx\}}\right]} ,\\
\mathbf{E}[|\rvx|^2|y] &= \frac{\mathbf{E} \left[|\rvx|^2 e^{-\gamma|\rvx|^2+2\alpha\Re\{\rvx\}-2\beta\Im\{\rvx\}}\right]}{\mathbf{E} \left[ e^{-\gamma|\rvx|^2+2\alpha\Re\{\rvx\}-2\beta\Im\{\rvx\}}\right]},
\end{align}
which, upon inserting back the expressions of $\alpha$, $\beta$ and $\gamma$, are exactly the conditions (\ref{eqn:gmi-maximizing-1st}) and (\ref{eqn:gmi-maximizing-2nd}) in Theorem \ref{thm:GMI-general-input}.

\section{Proof of Corollary \ref{cor:gap-mi-gmi}}
\label{sec:proof of corollary 1}
From the proof of Theorem \ref{thm:GMI-general-input}, we see that the considered optimal $(g, f)$ satisfies (\ref{eqn:gmi-opt-g-f});
\begin{figure*}
    \begin{align}
        \label{eqn:gmi-opt-g-f}
        I_{\mathrm{GMI}} = - \sum_{(x, y) \in \mathcal{X}\times \mathcal{Y}} p_\rvx(x) p_{\rvy|\rvx}(y|x) |g(y) - f(y) x|^2 - \sum_{y \in \mathcal{Y}} p_\rvy(y) \log \sum_{x' \in \mathcal{X}} p_\rvx(x') e^{- |g(y) - f(y) x'|^2}.
    \end{align}
\end{figure*}
on the other hand, the channel mutual information is given by
\begin{align}
    I(\rvx; \rvy) = \sum_{(x, y) \in \mathcal{X}\times \mathcal{Y}} p_{\rvx, \rvy}(x, y) \log \frac{p_{\rvx, \rvy}(x, y)}{p_\rvx(x) p_\rvy(y)}.
\end{align}
Evaluating $I(\rvx; \rvy) - I_\mathrm{GMI}$, with the auxiliary probability distribution $\tilde{p}_{\rvx|\rvy}$ in (\ref{eqn:auxiliary-p}) recognized, we obtain
\begin{align}
    I(\rvx; \rvy) - I_\mathrm{GMI} &= \sum_{y \in \mathcal{Y}} p_\rvy(y) \sum_{x \in \mathcal{X}} p_{\rvx|\rvy}(x|y) \log \frac{p_{\rvx|\rvy}(x|y)}{\tilde{p}_{\rvx|\rvy}(x|y)}\nonumber\\
    &= D(p_{\rvx|\rvy} \| \tilde{p}_{\rvx|\rvy} | p_\rvy).
\end{align}

\section{Proof of Theorem \ref{thm:GMI-opt-qpsk}}
\label{sec:proof of theorem 2}
We apply Theorem \ref{thm:GMI-general-input}. For QPSK, $|\rvx|^2 = P$ holds with probability one, and hence the condition (\ref{eqn:gmi-maximizing-2nd}) becomes trivial. We thus only need to inspect the condition (\ref{eqn:gmi-maximizing-1st}).

From the proof of Theorem \ref{thm:GMI-general-input}, we can use (\ref{eqn:alpha-beta-gamma-1st}), simplifying it, to rewrite (\ref{eqn:gmi-maximizing-1st}) as
\begin{align}
    \label{eqn:gmi-maximizing-1st-qpsk}
    \Re\{\mathbf{E}[\rvx|y]\} &= \sqrt{\frac{P}{2}} \cdot \frac{\sinh{\sqrt{2P} \alpha}}{\cosh{\sqrt{2P} \alpha}},\nonumber\\
    \Im\{\mathbf{E}[\rvx|y]\} &= -\sqrt{\frac{P}{2}} \cdot \frac{\sinh{\sqrt{2P} \beta}}{\cosh{\sqrt{2P} \beta}};
\end{align}
that is,
\begin{align}
    \label{eqn:gmi-maximizing-1st-qpsk-result}
    \alpha &= \frac{1}{\sqrt{2P}} \mathrm{artanh} \sqrt{\frac{2}{P}} \Re\{\mathbf{E}[\rvx|y]\},\nonumber\\
    \beta &= -\frac{1}{\sqrt{2P}} \mathrm{artanh} \sqrt{\frac{2}{P}} \Im\{\mathbf{E}[\rvx|y]\}.
\end{align}
Correspondingly, (\ref{eqn:obj-y-new}) becomes (\ref{eqn:obj-y-new-qpsk}). Taking the expectation of (\ref{eqn:obj-y-new-qpsk}) with respect to $p_\rvy$ as in (\ref{eqn:optimization-GMI-gf-decomposed}), we obtain (\ref{eqn:GMI-opt-qpsk}).
\begin{figure*}
\begin{align}
    \label{eqn:obj-y-new-qpsk}
    &\gamma P - 2 \alpha \Re\{ \mathbf{E}[\rvX|y]\} + 2 \beta \Im\{ \mathbf{E}[\rvX|y]\} + \log \mathbf{E} \left[ e^{-\gamma P + 2 \alpha \Re\{\rvx\} - 2\beta \Im\{\rvx\}}\right] = \sqrt{\frac{2}{P}} \Re\{\mathbf{\mathbf{E}}[\rvx| y]\} \mathrm{artanh}\sqrt{\frac{2}{P}} \Re\{\mathbf{\mathbf{E}}[\rvx| y]\} \nonumber\\ 
    & + \sqrt{\frac{2}{P}} \Im\{\mathbf{\mathbf{E}}[\rvx| y]\} \mathrm{artanh}\sqrt{\frac{2}{P}} \Im\{\mathbf{\mathbf{E}}[\rvx| y]\} - \log \mathbf{E}\left[
        e^{-\sqrt{\frac{2}{P}} \Re\{\rvx\} \mathrm{artanh}\sqrt{\frac{2}{P}} \Re\{\mathbf{\mathbf{E}}[\rvx| y]\} - \sqrt{\frac{2}{P}} \Im\{\rvx\} \mathrm{artanh}\sqrt{\frac{2}{P}} \Im\{\mathbf{\mathbf{E}}[\rvx| y]\}}
    \right].
\end{align}
\end{figure*}

For QPSK, the maximization of GMI only depends upon $\alpha$ and $\beta$. We can therefore choose $g$ and $f$ as long as the optimality condition (\ref{eqn:gmi-maximizing-1st-qpsk-result}) is satisfied. One such choice is
\begin{align}
    f(y) &= 1,\\
    g(y) &= \frac{1}{\sqrt{2P}}\mathrm{artanh}\sqrt{\frac{2}{P}} \Re\{\mathbf{E}[\rvX|y]\} \nonumber\\
    &\quad\quad + \frac{\jmath}{\sqrt{2P}}\mathrm{artanh}\sqrt{\frac{2}{P}}\Im\{\mathbf{E}[\rvX|y]\},
\end{align}
and the resulting optimal GNND is given by \eqref{eqn:GNND-qpsk-opt}.}

\end{document}